\documentclass[alpha-refs]{article}

\usepackage{amsthm}
\usepackage{amsmath}
\usepackage{amssymb}
\usepackage{natbib}
\usepackage{authblk}
\usepackage[margin=1in]{geometry}
\usepackage{caption}
\usepackage{tikz}

\theoremstyle{definition}

\title{Detecting regime transitions of the nocturnal and Polar near-surface temperature inversion}

\author[1]{Amandine Kaiser}
\author[2,3]{Davide Faranda}
\author[4]{Sebastian Krumscheid}
\author[5]{Danijel Belu\u{s}i\'{c}}
\author[1]{Nikki Vercauteren}

\affil[1]{FB Mathematik und Informatik, Freie Universit{\"a}t Berlin, 14195 Berlin, Germany}
\affil[2]{LSCE-IPSL, CEA Saclay l'Orme des Merisiers, CNRS UMR 8212 
CEA-CNRS-UVSQ, Universit\'e Paris-Saclay, 91191 Gif-sur-Yvette, France}
\affil[3]{London Mathematical Laboratory, London, United Kingdom}
\affil[4]{Department of Mathematics, RWTH Aachen, Aachen, Germany}
\affil[5]{SMHI, Norrk\"oping, Sweden}

\begin{document}

\maketitle

\begin{abstract}Many natural systems undergo critical transitions,
  i.e.~sudden shifts from one dynamical regime to another. In the
  climate system, the atmospheric boundary layer can experience sudden
  transitions between fully turbulent states and quiescent,
  quasi-laminar states. Such rapid transitions are observed in Polar
  regions or at night when the atmospheric boundary layer is stably
  stratified, and they have important consequences in the strength of
  mixing with the higher levels of the atmosphere. To analyze the
  stable boundary layer, many approaches rely on the identification of
  regimes that are commonly denoted as weakly and very stable
  regimes. Detecting transitions between the regimes is crucial for
  modeling purposes. \\
  In this work a combination of methods from dynamical systems and
  statistical modeling is applied to study these regime transitions
  and to develop an early-warning signal that can be applied to non-stationary field data. The presented metric aims at detecting nearing transitions by statistically quantifying the deviation from the dynamics expected when the system is close to a stable equilibrium. An idealized stochastic model of near-surface inversions is used to evaluate the potential of the metric as an indicator of regime transitions. In this stochastic system, small-scale perturbations can be amplified due to the nonlinearity, resulting in transitions between two possible equilibria of the temperature inversion. The simulations show such noise-induced regime transitions, successfully identified by the indicator. The indicator is further applied to time series data from nocturnal and Polar meteorological measurements.
\end{abstract}


%

\section{Introduction}
The atmospheric boundary layer (ABL) is the lowest part of the
atmosphere that is directly influenced by the Earth's surface and
across which turbulent exchanges of momentum, heat and matter between
the surface and the free atmosphere occur. During daytime, surface
warming leads to an unstable or convective boundary layer. During
clear-sky nights, radiative cooling leads to a surface that is cooler
than the air aloft and the ABL becomes stably stratified. The stable
stratification can also arise when warm air is advected over a colder
surface, which is a frequent event in Polar regions. Turbulence in the
resulting stable boundary layer (SBL) is subject to buoyant damping
and is only maintained through mechanical production of turbulent
kinetic energy (TKE). Understanding and modeling the SBL is essential
for regional and global atmospheric models, yet there are many
well-documented challenges to simulate stably stratified atmospheric
flows \citep{Sandu:2013ia, Holtslag:2013iy, LeMone:2018gx}. One of the
challenges is to develop an accurate understanding and representation
of distinct regimes of the SBL and transitions between them \citep{Baas:2017fl}.

Numerous observational and modeling studies show that the SBL can be
classified, to a first approximation, in a weakly stable regime in which
turbulence is continuous, and a very stable regime with patchy and
intermittent turbulence, requiring a different modeling approach
\citep{Mahrt:2014wc}. The weakly stable regime typically occurs when
cloud cover limits nocturnal radiative cooling at the land surface, or
with strong winds associated to wind shear that produces enough TKE to
sustain turbulence. The vertical mixing is therefore
maintained and a well-defined boundary layer usually exists in which
turbulent quantities decrease upwards from the surface layer following
the classical model of Monin-Obukhov similarity theory and related
existing concepts \citep{Mahrt:2014wc}. The associated temperature
stratification, or temperature inversion, is weak. The strongly stable
regime occurs with strong stratification and weak winds and does not
follow the traditional concept of a boundary layer. Transitions from
weakly stable to strongly stable regimes are caused by a strong net
radiative cooling at the surface which increases the inversion
strength and eventually leads to suppressed vertical exchanges, unless
winds are strong enough to maintain turbulence
\citep{vandeWiel:2007ev}. The reduced vertical mixing results in a
decoupling from the surface, such that similarity theory breaks down
\citep{Acevedo:2015iw}. Intermittent bursts tend to be responsible for
most of the turbulent transport \citep{Acevedo:2006bl,
  Vercauteren:2016kx}. Such bursts alter the temperature inversion and
can sometimes drive transitions from strongly to weakly stable
boundary layers.

Transitions between the different SBL regimes have been found by
modeling studies to be dynamically unpredictable. Based on a
numerical model representing the exchanges between the surface and the
SBL using a very simple two-layer scheme, \cite{McNider:1995jc} showed
the existence of bi-stable equilibria of the system, which can
thus transition between very different states under the influence of
random perturbations. Interacting nonlinear processes that lead to
this bi-stability partly involve thermal processes at the land
surface, as was highlighted following the hypothesis that continuous
turbulence requires the turbulence heat flux to balance the surface
energy demand resulting from radiative cooling
\citep{vandeWiel:2007ev, vandeWiel:2012jy,
  vandeWiel:2017ju}. According to this maximum sustainable heat flux
hypothesis, a radiative heat loss that is stronger than the
maximum turbulent heat flux that can be supported by the flow with a
given wind profile will lead to the cessation of turbulence
\citep{vandeWiel:2012ji} and thus to a regime transition. This concept
is used by \cite{vanHooijdonk:2015gh} to show that the shear over a
layer of certain thickness can predict SBL regimes when sufficient
averaging of data is considered. Based on observations,
\cite{Sun:2012eo} identify a height and site-dependent wind speed
threshold that triggers a transition between a regime in which
turbulence increases slowly with increasing wind speed from a regime
where turbulence increases rapidly with the wind speed. The change of
the relationship between the turbulence and the mean wind speed occurs
abruptly at the transition.  \cite{Sun:2016ep} attribute this
difference to the turbulent energy partitioning between turbulent
kinetic energy and turbulent potential energy (TPE): in the very
stable regime, shear-induced turbulence will have to enhance the TPE
in order to counter the stable stratification before enhancing the
TKE.

The combined importance of the wind speed and of the surface thermal
processes has also been evidenced by numerical studies using idealized
single-column models of the atmosphere. Single-column models with a
first-order turbulence closure scheme \citep{Baas:2017fl, Baas:2019cl,
  Holdsworth:2019iu} or a second-order closure scheme
\citep{Maroneze:2019cc} are able to representatively simulate transitions from weakly
to strongly-stable regimes. Yet, direct numerical
simulations show that transitions from strongly to weakly-stable
regimes can occur following a localized, random perturbation of the
flow \citep{Donda:2015kh}. Field studies have also highlighted
examples of transitions induced by small-scale perturbations of the
flow \citep{Sun:2012eo}. In fact, a statistical classification scheme
introduced by \cite{Vercauteren:2015fq} shows that the SBL flow
transitions between periods of strong and weak influence of
small-scale, non-turbulent flow motions on TKE production in the
SBL. Such submeso-scale fluctuations of the flow (e.g. induced by various kind
of surface heterogeneity)
are typically not
represented in models but are important in strongly stable regimes
\citep{Vercauteren:2019fw}, and may trigger regime
transitions. Stochastic modeling approaches are a promising framework to
analyze their impact on regime transitions.  A related statistical
classification of the Reynolds-averaged boundary-layer states
introduced by \cite{Monahan:2015fu} highlighted that regime
transitions are a common feature of SBL dynamics around the globe
\citep{Abraham:2019a}. Regime transitions typically take an abrupt
character \citep{Baas:2019cl, Acevedo:2019fg}. Predicting the
transition point remains a challenge
\citep{vanHooijdonk:2016bw}.

Abrupt or critical transitions are ubiquitous in complex natural and
social systems. The concept of critical transition is formally defined
in dynamical systems theory and relates to the notion of bifurcation
\citep{kuznetsov2013elements}. When the dynamics is controlled by a
system of equations depending on an external parameter (often called
forcing), the stability of the equilibrium solutions can change
abruptly and this is also reflected on macroscopic observables of the
system. Sometimes, one can have early-warning signals of a transition
because the systems experience some influences of the bifurcated state
before actually reaching it. The motion of a particle undergoing
random fluctuations in an asymmetric double-well energy potential $V$
is a minimal system to detect early warnings, in which each well or
local minimum of the energy potential corresponds to a stable
equilibrium state of the system. For a small fixed level of noise, the
control parameter is $\Delta V$, the depth of the well in which the
particle is located, leading to an energy barrier that the particle
has to overcome in order to transition to the second stable
equilibrium. If $\Delta V$ is large, the distribution of the positions
of the particle will be quasi Gaussian and the autocorrelation
function of the position of the particle will have an exponential
decay. Conversely, if $\Delta V$ is reduced when the particle
approaches the bifurcation point, then the particle position's
distribution starts to “feel” the effect of the other state and the
distribution will be skewed towards the new state. Similarly, the
excursions from the equilibrium position will become larger,
increasing the autocorrelation time. Early-warning signals can then,
e.g., be defined based on the changes of the autocorrelation time.

These first early-warning signs have been successfully applied to
several systems with excellent results
\citep{scheffer2009early,scheffer2012anticipating}, including the
present context of SBL regime transitions
\citep{vanHooijdonk:2016bw}. However, sometimes, transitions can
happen without detectable early-warnings
\citep{hastings2010regime}. The main limitation of early-warning
signals based on the increase of autocorrelation is that their
activation does not always correspond to a bifurcation. Indeed, if a
single well potential widens, as it can occur in non-stationary
systems, the distribution of a particle's position experiences the
same increase in skewness and autocorrelation function without the
need of approaching a bifurcation
\citep{lenton2012early,Faranda_2014}. In our context of SBL flows,
non-stationarity of the energy potential governing the dynamics can be
due to changes in the mean wind speed or cloud cover, for example. For
these reasons, \citep{Faranda_2014} have introduced a new class of
early-warning indicators based on defining a
distance from the dynamics expected from a particle evolving in a
single-well potential. The suggested indicator statistically
quantifies the dynamical stability of the observables and was already
used by \cite{Nevo2017StabInd} to show that strongly stable flow
regimes are dynamically unstable and may require high-order turbulence
closure schemes to represent the dynamics. Alternative new early
warnings are based on the combination of statistical properties of
observables when approaching the bifurcation
\citep{chen2012detecting}.

In the present analysis, we investigate if the early-warning indicator
introduced by \cite{Faranda_2014} can be used to detect nearing
transitions between SBL flow regimes, based on both simulated data and
field measurements. We show that the conceptual model that was
recently suggested by \cite{vandeWiel:2017ju} to understand SBL regime
transitions in terms of thermal coupling of the land surface is
equivalent to a dynamical system representing the evolution of the
temperature inversion evolving in a double-well energy potential. We
extend this conceptual model to a stochastic model where added noise
represents the effect of natural fluctuations of the temperature
inversion's rate of change. The resilience of equilibria of the
non-random model to perturbations as well as the bifurcation points
are known analytically (as was discussed in \cite{vandeWiel:2017ju}),
and we thus use the simulated data to test our
indicator. Additionally, the indicator relies on calculating
statistical properties of the data with a moving window approach and
is sensitive to the choice of the window length. We suggest two
complementary, data-driven but physically justified approaches to
define an appropriate window length for which results can be
trusted. Finally, the indicator is applied to nocturnal temperature
inversion data from a site in Dumosa, Australia as well as from
temperature inversion data from Dome C, Antartica.

\section{Analyzing the dynamical stability of stable boundary layer regimes}\label{sec:AnalysesDynStab}
The goal of our study is to investigate if a statistical early-warning
indicator of regime transitions can be successfully used to detect nearing
regime transitions in the SBL. In section
\ref{subseq:ModelDescription}, the conceptual model introduced by
\cite{vandeWiel:2017ju} to study regime transitions will be
introduced, along with its dynamical stability properties. In section \ref{subseq:SDEModel}, the model is extended to a stochastic model in which noise represents fluctuations in the dynamics of the near-surface temperature inversion. In section
\ref{subseq:StatIndicator}, we present a statistical indicator that
was introduced in \cite{Faranda_2014} and applied to SBL
turbulence data in \cite{Nevo2017StabInd} to estimate the dynamical
equilibrium properties of time series, based on a combination of
dynamical systems concepts and stochastic processes tools. The
conceptual model describes the evolution of the near-surface
temperature inversion and is used to produce time series of controlled
data for which the theoretical equilibrium properties are known.

\subsection{Model description and linear stability analysis}\label{subseq:ModelDescription}
A conceptual model was introduced by \cite{vandeWiel:2017ju} to study
regime transitions of near-surface temperature inversions in the
nocturnal and polar atmospheric boundary layer. The authors were able
to determine a connection between the dynamical stability of the
temperature inversion and the ambient wind speed $U$ through their
model and measurements. Mathematically speaking, the model is a
dynamical system represented by a first order ordinary differential
equation, abbreviated ODE, which describes the time evolution of the
difference between the temperature at a
reference height $T_r$ and the surface temperature $T_s$. Although the equilibrium properties of the
system and the dynamical stability properties (i.e. the resilience to
perturbations) of all equilibria states were thoroughly discussed in
\cite{vandeWiel:2017ju}, for the sake of completeness we briefly introduce the
model and summarize the linear stability analysis of equilibrium
points of the resulting ODE for different values of a bifurcation
parameter. The bifurcation parameter is related to the ambient wind speed.
\newline \newline Assuming that the wind speed and temperature are
constant at a given height $z_r$, the following equation describes the
evolution of the near-surface inversion strength, based on a simple
energy balance at the ground surface:
\begin{align}
	c_v\frac{d\Delta T}{dt}=Q_n-G-H. 
\end{align}
In this energy balance model, $c_v$ is the heat capacity of the soil, $\Delta T=T_r-T_s$ is the inversion strength between the temperature at height $z_r$ and at the surface $z_s$, $Q_n$ is the net long wave radiative flux (an energy loss at the surface that will be set as a constant), $G$ is the soil heat flux (an energy storage term that will be parameterized as a linear term), and $H$ is the turbulent sensible heat flux (a non-linear energy transport term that will be parameterized in the following).

After parameterizing the fluxes, the model has the form:
\begin{align} \label{Eq:EBM}
	c_v\frac{d\Delta T}{dt}=Q_i-\lambda \Delta T-\rho c_p c_D U \Delta T f(R_b),
\end{align}
in which $Q_i$ is the isothermal net radiation, $\lambda$ is a lumped parameter representing all feedbacks from soil heat conduction and radiative cooling as a net linear effect, $\rho$ is the density of air at constant pressure, $c_p$ is the heat capacity of air at constant pressure, $c_D=(\frac{\kappa}{ln(z_r/z_0)})^2$ is the neutral drag coefficient with $\kappa \approx 0.4$ the von K\'arm\'an constant, $z_0$ the roughness length and $z_r$ the reference height, $U$ is the wind speed at height $z_r$, $R_b=z_r(\frac{g}{T_r})\frac{\Delta T}{U^2}$ is the bulk Richardson number, and $f(R_b)$ is the stability function used in Monin-Obukhov similarity theory.

The lumped parameter $\lambda$ corresponds to a linear term in the model as the soil is assumed to respond linearly to the temperature inversion. Moreover, $\Delta T \cdot f(R_b)$ is a non-linear term due to the non-linear dependence of turbulent diffusion on the vertical temperature gradient.
\newline \newline
Following \cite{vandeWiel:2017ju}, instead of analyzing the dynamical stability of the energy-balance model (\ref{Eq:EBM}) itself, we will present the linear stability analysis of a simplified system that has a similar mathematical structure but is mathematically convenient to analyze. Using a cutoff, linear form for the stability function, i.e. $f(x)=1-x$ and $f(x) = 0$ for $x>1$, the simplified model is
\begin{equation}
  	\frac{dx(t)}{dt} = g\bigl(x(t)\bigr)\;,\quad\text{where}\quad
        g(x) = \begin{cases}
          Q_i-\lambda x-Cx(1-x) & \text{for } x\leq 1\\  \label{eq:ODEvandeWiel}
          Q_i-\lambda x & \text{for } x>1
        \end{cases}
\end{equation}
and $x(t_0)=x_0$. Here, up to dimensional constants, $x$ represents
$\Delta T$. The parameter $C$ will be treated as a bifurcation
parameter for this simplified system.  Similar types of stability
functions are typically used in numerical weather prediction tools,
and the cutoff form facilitates the mathematical analysis of the
model. Note that to be consistent with the original model, the
stability function should include a dependence on both the temperature
and the wind speed via $R_b$. Removing this dependence as it is done
here changes some of the nonlinearity, however it makes the
mathematical analysis very simple and the qualitative behavior of the
system is similar to the original system - see
\cite{vandeWiel:2017ju}, their Figure 8 and 10. In that sense, the
model loses some physical significance for mathematical convenience,
but the qualitative nonlinear feedback processes are maintained. For a
deeper discussion of the model, its simplifications and the model
parameters, the reader is referred to its thorough presentation by
\cite{vandeWiel:2017ju}.

\begin{figure}[hbt]
\centering
	\includegraphics[width=0.5 \textwidth]{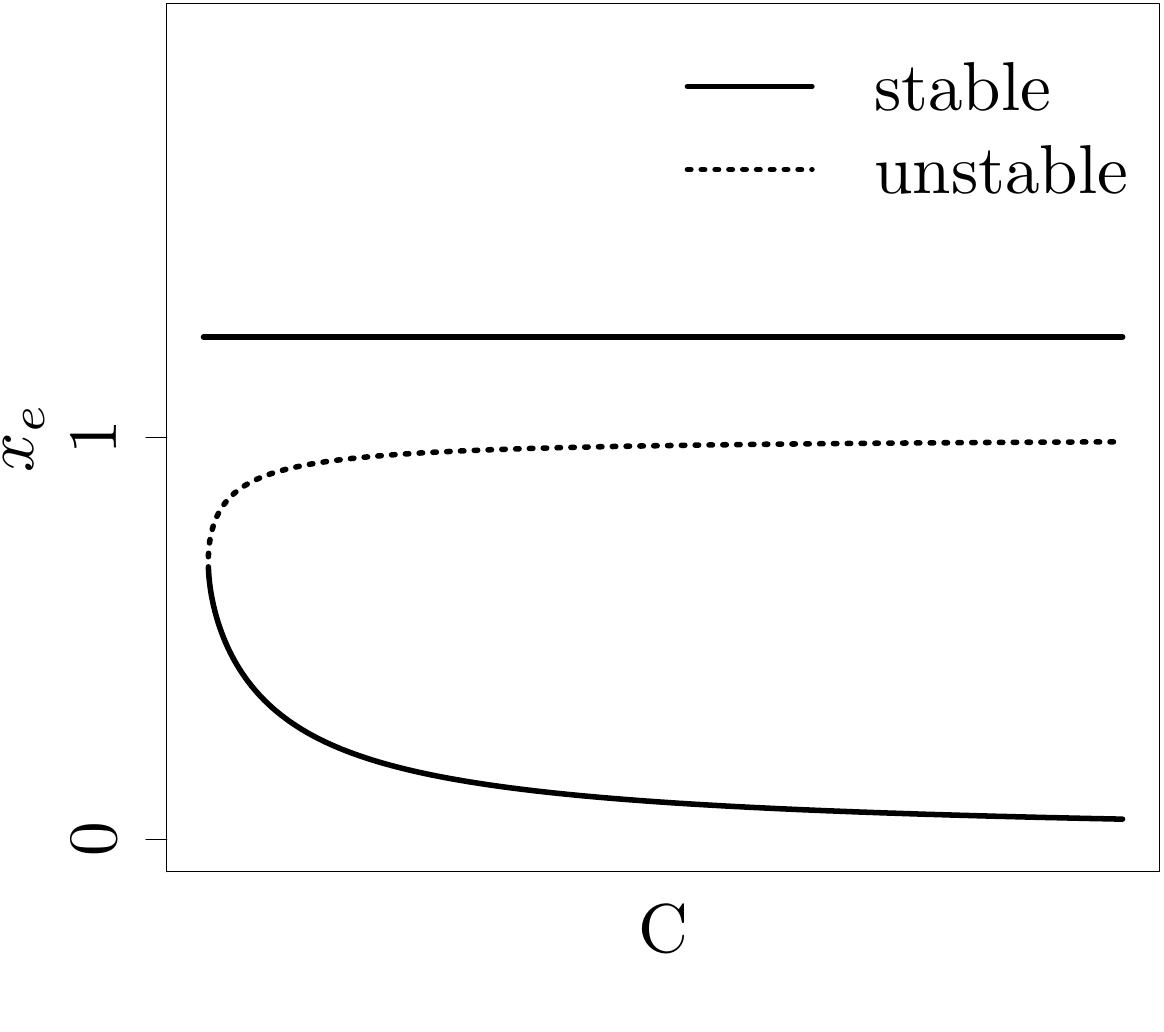}
	\caption{Bifurcation plot for simplified model. The lowest, middle and upper branches correspond respectively to the equilibria $x_{e_1}$, $x_{e_2}$ and $x_{e_3}$}\label{fig:fig_bifurcation}	
\end{figure}

For fixed and physically meaningful values of $Q_i$ and $\lambda$,
equation \eqref{eq:ODEvandeWiel} can have either one, two, or three
possible equilibrium solutions depending on the fixed values (see
illustration in \cite{vandeWiel:2017ju}, Figure 10-12, and related
discussion for more details). The equilibrium solutions will be
functions of the parameter $C$, which we will consider as a
bifurcation parameter in the following. Physically, the case of strong
thermal coupling between the surface and the atmosphere, corresponding
to a large value of $\lambda$, results in one unique equilibrium
solution whose value depends on $C$. In \cite{vandeWiel:2017ju}, it is
hypothesized that such a case is representative of a grass site such
as Cabauw, the Netherlands. The solution is linearly stable to
perturbations, i.e. linear stability analysis shows that perturbed
solutions are attracted back to the equilibrium. The case of no
coupling ($\lambda$=0) leads to two equilibrium solutions, one of
which is linearly stable and the other unstable to perturbations
(i.e. perturbed solutions are repelled by the equilibrium). A weak
coupling strength, with an intermediate value of $\lambda$ that could
be representative of a snow surface or another thermally insulated
ground surface, results in three possible equilibrium solutions. The
two extreme solutions are stable to perturbations, while the middle
equilibrium solution is unstable. Perturbed solutions around the
middle equilibrium will thus be attracted either by the upper or the
lower equilibrium. Plotting those three equilibrium solutions as a
function of the bifurcation parameter $C$ results in a back-folded
curve which is qualitatively similar to observations of the
temperature inversion shown as a function of wind speed at Dome C,
Antartica; see \cite{Vignon:2017bf}. The bifurcation diagram is shown
in Figure \ref{fig:fig_bifurcation} for parameter values such that
$\lambda>0$ and $Q_i > \lambda$, resulting in the case with three
possible equilibrium solutions. 
By convention, the unstable
equilibrium branch is denoted by a dashed line. In the following, we
will analyze transitions between the two stable equilibria. If the
system undergoes random perturbations in this bi-stable context, a
perturbation could drive the system sufficiently far from a stable
equilibrium state so that it comes near the unstable equilibrium and
finally gets attracted by the second equilibrium. The three possible
equilibria are denoted as $x_{e_1}$, $x_{e_2}$ and $x_{e_3}$. In order
to study such regime transitions induced by random perturbations, the
conceptual model is extended with a noise term in the following
section.

\subsection{Extending the conceptual model by
  randomization} \label{subseq:SDEModel} 
\begin{figure}[hbt]
\centering
	\noindent
	\includegraphics[width=0.9\textwidth]{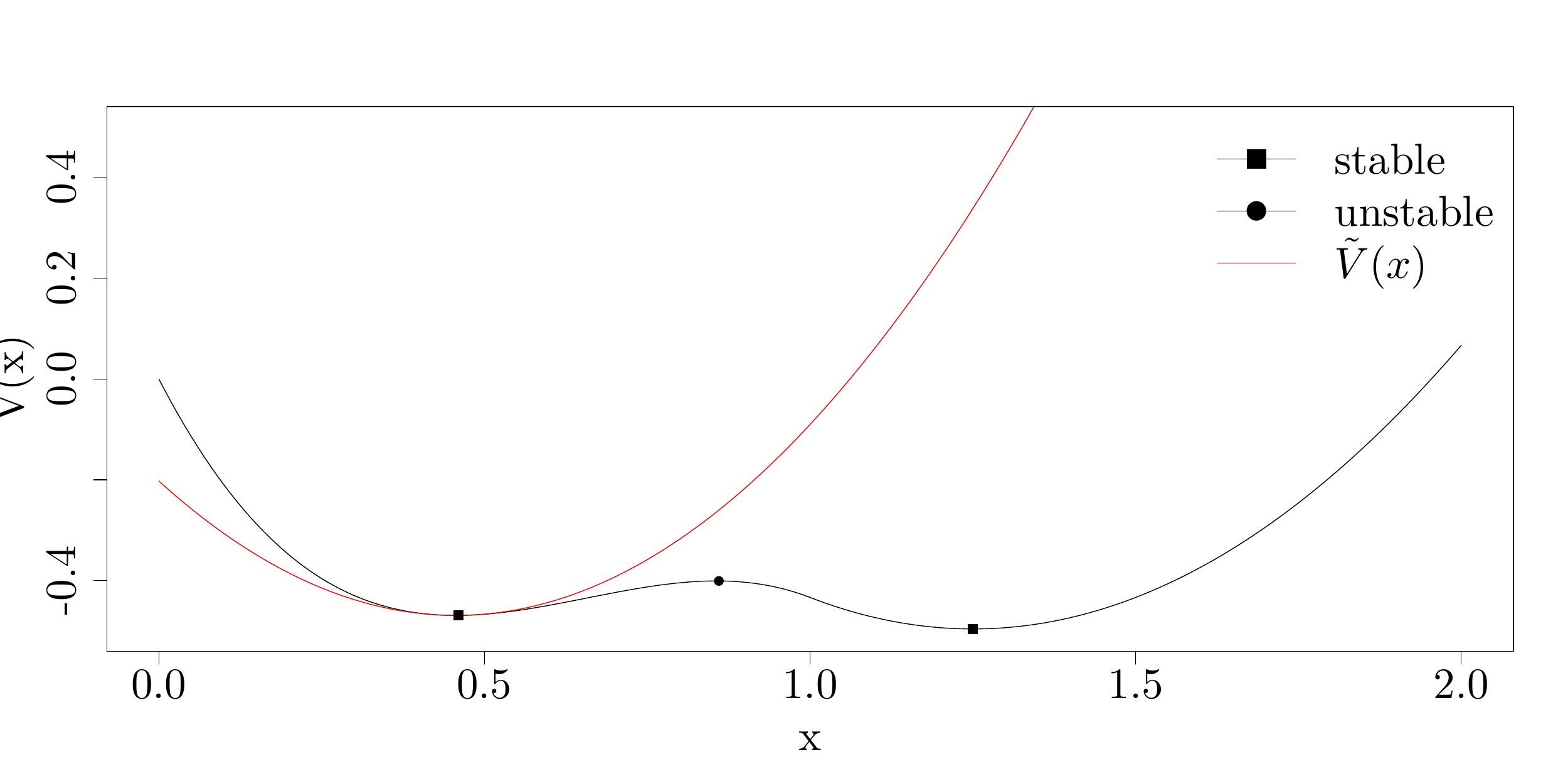}
	\caption{Example of a potential $V(x)$ (eq. \ref{eq:PotVanDeWiel}) and its local approximation through the quadratic potential $\tilde{V}(x)$ (see text for details).}\label{fig:potential}
\end{figure}

The conceptual model
\eqref{eq:ODEvandeWiel} can be equivalently written in terms of a
gradient system, in which the temperature inversion represented here
by $x$ evolves according to the influence of an underlying potential
$V(x)$. The randomized model to be introduced will be based on this
gradient structure. Specifically, the initial value problem
(\ref{eq:ODEvandeWiel}) can be written as
\begin{equation*}
  \frac{dx}{dt}=-\frac{dV}{dx}\;,\quad x(t_0) = x_0\;, 
\end{equation*}
where it is easy to see that the potential is given by
\begin{equation}
V(x)=
\begin{cases}
\frac{1}{2} x^2 (\lambda+C)-\frac{C}{3}x^3-Q_i x & \text{for } x\leq 1\;,\\
\frac{1}{2}\lambda x^2-Q_i x+\frac{1}{6}C & \text{for } x>1\;.
\end{cases}\label{eq:PotVanDeWiel}
\end{equation}
The linear stability analysis discussed in the previous section can
thus be understood in the sense that the temperature
inversion 
$x$ equilibrates at a local minimum of a potential $V$. That is, an
equilibrium point $x_e$ satisfies $V'(x_e) = 0$.  Figure
\ref{fig:potential} sketches the form of the potential with the
exemplary parameter values $\lambda=2, Q_i=2.5, C=6.4$. Note that
$V(x)$ is a double-well potential in that case where each local
minimum corresponds to one of the stable equilibrium points $x_{e_1}$
and $x_{e_3}$, while the local maximum corresponds to the system's
unstable equilibrium $x_{e_2}$.

While the conceptual model \eqref{eq:ODEvandeWiel} has proven very
insightful to explain observed sharp transitions in temperature
inversions, it only allows for regime transitions when drastic changes
in the model \emph{parameters} (i.e., bifurcations) occur. That is, the model
is overly idealized and in reality one can expect regime transitions
to also take place due to small natural fluctuations of the
temperature inversion itself in certain cases, e.g., when the potential
barrier separating the two local minima and corresponding stable
equilibria is shallow. Therefore we will consider an appropriate
randomized variant of the model. Specifically, we consider the
stochastic differential equation (SDE) model
\begin{equation}
  dx = -\frac{dV(x)}{dx}\,dt + \sigma \,dB\;,\quad x(t_0) = x_0\;, \label{eq:ODEvandeWiel:SDE}
\end{equation}
to account for small random perturbations to the temperature
inversion's rate of change. Here, $B$ denotes a standard Brownian
motion (i.e., a stochastic process) and $\sigma>0$ scales the
intensity of the fluctuations, while the potential $V$ is as in
\eqref{eq:PotVanDeWiel}. As the randomized dynamics is characterized
by the same potential, also the equilibrium points of the non-random
model \eqref{eq:ODEvandeWiel} will describe the dominant effects of
the randomized model's dynamics. However, due to the presence of the
noise, the stable equilibria of the non-random model
\eqref{eq:ODEvandeWiel} are not limiting points for the stochastic
counter-part in \eqref{eq:ODEvandeWiel:SDE}, in the sense that the
temperature inversion may still fluctuate after reaching a stable
equilibrium. The reason is that in a context of two stable equilibria
(i.e. for parameter values such that the model \eqref{eq:ODEvandeWiel}
exhibits two stable equilibria, denoted earlier as the thermally
weakly-coupled state), the random perturbations can trigger
transitions from one stable equilibrium to another one. We will
therefore refer to the formerly stable states as: metastable. Note
that depending on the coupling strength and noise intensity, the
likelihood of regime transitions can change drastically and the system
may or may not exhibit metastable states. The type of noise (additive
or multiplicative for example, or noise with a Levy distribution) will
also affect regime transitions. 
In our subsequent simulations and analyses, we will focus on the case
of two metastable states with additive noise and leave other cases for
future research.

The effect of these random perturbations to a metastable equilibrium
point $x_e$ can be understood through a localized approximation of the
original dynamics. More precisely, consider a second-order Taylor
approximation of the potential around an equilibrium point $x_{e}$,
yielding the quadratic approximate potential $\tilde{V}$:
\begin{equation*}
  V(x) \approx \tilde{V}(x) := V(x_e) +  \frac{1}{2}{\left.\frac{d^2V}{dx^2}(x)\right\vert}_{x=x_e}{(x-x_e)}^2\;.
\end{equation*}
For the same parameter values that were used to plot the original
potential in Figure \ref{fig:potential}, the red line in the same
figure shows the approximate quadratic potential around the
equilibrium value $x_{e_1}$.  
Using the locally quadratic potential,
we can thus define a locally approximate dynamics for the temperature
inversion by replacing $V$ in \eqref{eq:ODEvandeWiel:SDE} by
$\tilde{V}$, resulting in
\begin{equation*}
  dX = -k(X-x_e)\,dt + \sigma \,dB\;,\quad X(t_0) = x_e\;,
\end{equation*}
where $k:=\frac{d^2V}{dx^2}(x)|_{x=x_e} \in\mathbb{R}$ and $X$ is introduced to describe the approximate dynamics of the former $x$. This
approximate dynamics is an example of the well-studied
Ornstein--Uhlenbeck process and it provides an accurate description of
the full dynamics in the neighborhood to the equilibrium point $x_e$.
Discretizing the Ornstein--Uhlenbeck process $X$ using the
Euler--Maruyama scheme with a step-size $\Delta t:=\frac{T}{L}$ for
some positive integer $L$ we furthermore find that the process at
discrete times $t\in\{1,\dots, L\}$ approximately satisfies the difference equation
\begin{equation*}
  X_{t} = X_{t-1} - k(X_{t-1}-x_e) \Delta t + \sigma\Bigl(B(t\Delta t)-B\bigl((t-1)\Delta t\bigr)\Bigr)\;,\quad X_0 = e_e\;,
\end{equation*}
in the sense that $X_t\approx X(t\Delta t)$.
By defining $\mu:=kx_{e} \in \mathbb{R}$, $\phi:=(1-k\Delta t)$ and
$w_t:=\sigma\bigl[B(t\Delta t)-B\bigl((t-1)\Delta t\bigr)\bigr]$ this can be written as
\begin{equation*}
  X_t=\mu+\phi X_{t-1}+w_t\;,
\end{equation*}
which is a so-called autoregressive model of order one, denoted AR(1),
thanks to the properties of the (scaled) Brownian increments
$w_t$. Consequently, we see that the discretized Ornstein--Uhlenbeck
process can be accurately approximated by an AR(1) process. This derivation can also be found in \cite{Thomson:1987ge}. Combining
this with the observation that the Ornstein--Uhlenbeck process offered
an accurate approximation to the original dynamics in the vicinity of
a stable equilibrium, we can thus conclude that the local dynamics in
the neighborhood of a metastable state can be approximately described
by an AR(1) process.

\subsection{Statistical indicator for the dynamical stability of time series}\label{subseq:StatIndicator}
In section \ref{subseq:ModelDescription} we discussed the simplified
model by \citet{vandeWiel:2017ju} which was developed to understand
regime transitions in near-surface temperature inversions. This model
provides a hypothesis that explains the existence of two possible
equilibria of the temperature inversion for a given wind speed. In
agreement with the randomized conceptual model introduced in the
previous section, we say that a system exhibiting at least two
metastable equilibria is called metastable. In this section the goal
is to describe a methodology for statistically detecting critical
transitions based on time series data. For the detection we apply an
indicator for the dynamical stability (i.e. the resilience to
perturbations) of time series, which was defined by
\cite{Faranda_2014} and applied to SBL turbulence data in
\cite{Nevo2017StabInd}. The indicator uses a combination of methods
from dynamical systems and from statistical modeling. In its
definition, deviations from AR(1) processes in the space of so-called
autoregressive-moving-average (ARMA) models are used to quantify the
dynamical stability of a time series. A time series ${x_t}$, $t\in\mathbb{Z}$, is an
ARMA(p,q) process if it is stationary and can be written as
\begin{equation}
x_t= \nu + \sum_{i=1}^{p}\phi_i x_{t-i}+w_t+\sum_{j=1}^{q}\theta_j w_{t-j}\;, \label{eq:ARMA(p,q)}
\end{equation}
with constant $\nu$, coefficients $\phi_p$, $\theta_q$ and $\{w_t\}$ being
white noise with positive variance $\sigma^2$. The coefficients $\phi_p$ and $\theta_q$ additionally have to satisfy some constraints, see \cite{BrockwellPeterJ2016Itts}. Notice that AR(1) =
ARMA(1,0). Intuitively the parameters $p$ and $q$ are related to the
memory lag of the process. The longer the system takes to return to
the equilibrium after a perturbation, the more memory we expect to
observe in the process. Examples of simple systems along with their ARMA(p,q) characteristics can be found in \cite{Faranda_2014}.

In section \ref{subseq:SDEModel}, it was shown that the dynamics when
the system is close to a stable equilibrium can be approximated by a
AR(1) process. We will assume that far from the transition from one
dynamical regime to another, the time series of a generic physical
observable can be described by an ARMA(p,q) model with a reasonably
low number of $p,q$ parameters and coefficients.  Indeed far from a
transition, the system will tend to remain around an equilibrium
despite random perturbations, and excursions from the equilibrium are
short. The idea behind the modeling assumption is that ARMA processes
are an important parametric family of stationary time series
\citep{BrockwellPeterJ2016Itts}. Their importance is due to their
flexibility and their capacity to describe many features of stationary
time series. Thereby, choosing ARMA(p,q) processes for modeling the
dynamics away from a stable state is a reasonable Ansatz. Close to a
transition, the resilience of the system to perturbations is weak and
longer excursions from the equilibrium occur. The statistical
properties (such as the shape and/or the persistence of the
autocorrelation function) of the system change drastically, leading to
an expected increase of the value $p+q$ \citep{Faranda_2014}. Based on
this, we use ARMA(p,q) models in the following to analyze the
stability of a dynamical system. The dynamical stability indicator
which will be defined next will be used to obtain
indicators for detecting the system's proximity to
  a regime transition.

  In order to quantify the local stability of a time series, we first
  slice the time series $x_t$ with a moving time window of fixed
  length $\tau$. In other words, we obtain subsequences
  $\{x_1,\dots,x_{\tau}\},
  \{x_2,\dots,x_{\tau+1}\},\dots,\{x_{t-\tau+1},\dots,x_t\}$ of the
  original time series that overlap. By slicing the original time
  series we obtain a sequence of shorter time series for which it is
  reasonable to suppose that they are amenable to ARMA modeling. In
  detail, we assume that the subsequences are realizations of linear
  processes with Gaussian white noise which then implies that the
  process is stationary. We then fit an ARMA(p,q) model for every
  possible value of $(p,q)$, with $p\leq p_{max}$ and $q\leq q_{max}$,
  to these subsequences, where $p_{max}$ and $q_{max}$ are predefined
  thresholds. Afterwards we choose the best fitting ARMA(p,q) model by
  choosing the one with the minimal Bayesian information criterion,
  BIC, \citep{Schwarz1978BIC} which is a commonly used and
  well-studied tool in statistical model selection. Assuming that we
  have the maximum likelihood estimator
  $\hat{\beta} := (\hat\nu,\hat\phi_1,\dots,
  \hat\phi_p,\hat\theta_1,\dots, \hat\theta_q)$ of the fitted
  ARMA(p,q) model (which can be obtained using a so-called innovation
  algorithm, as it is, for example, implemented in the "forecast" R
  package \citep{forecastRpackage} which is used for the analyses),
  the BIC is formally defined as
\begin{equation}\label{Eq:BIC}
  \text{BIC}(p,q) =  -2\ln L(\hat{\beta}) + \ln(\tau)(p+q+1)\;,
\end{equation}
where $L(\hat{\beta})$ denotes the associated likelihood function
evaluated at the maximum likelihood estimator $\hat{\beta}$.  The
second term introduces a penalty for high-order models (i.e., those
that contain more parameters) to avoid over-fitting.

We reiterate that when the system is close to a metastable state, its
dynamics can be well approximated by an AR(1) process. When the system
is approaching an unstable point separating multiple basins of
attraction, the approximation no longer holds as the underlying
potential cannot be approximated by a quadratic potential anymore. The
change in the shape of the potential introduces new correlations in
the time series, resulting in higher-order ARMA terms when fitting
such a model to data.

The definition of the stability indicator is based on this
observation, in the sense that it assumes that the dynamics near a
metastable state can be modeled by an ARMA(1,0) or
AR(1)-process. Specifically, the stability indicator is defined as
\begin{align}
\Upsilon(p,q;\tau)=1-\exp\left(\frac{-|\text{BIC}(p,q)-\text{BIC}(1,0)|}{\tau}\right).
\end{align}
For a stable state, the most likely statistical model is an AR(1)
process and one expects that $\Upsilon=0$. The indicator $\Upsilon$
gives a normalized distance between the stable state ($\Upsilon=0$)
and the state in which the system is. The limit $\Upsilon\rightarrow1$
corresponds to a most likely statistical model of high order and probably to a nearing transition. While a formal proof of this
statement is still missing, tests performed for systems of increasing
complexity in \cite{Nevo2017StabInd} showed promising results where
the indicator correctly identified changes in the stability of
metastable states. To simplify the notation we drop the dependence of
$\Upsilon$ on $p,q$ and $\tau$ in the following discussion.

The reliability of $\Upsilon$ highly depends on the choice of $\tau$,
the window length (which we will consider in number of discrete
observations in the following), and it relates to the characteristic
timescales of the dynamics. Intuitively, the window length, when
converted to its equivalent physical duration (i.e. the number of
discrete observations multiplied by the discrete sampling time), has
to be shorter than the residence timescale in one basin of attraction
(i.e. the time spent in the neighborhood of an equilibrium before
transitioning to another one) in order to satisfy the local
stationarity, but large enough so that statistical model estimation is
reliable. In winter at Dome C where the Polar winter results in a near
absence of daily cycle, no preferred timescale of residence around an
equilibrium of the temperature inversion was observed (an equilibrium
can remain for several days), however the transition between two
equilibria was observed to take place over a timescale of the order of
10 hours \citep{Baas:2019cl}. For nocturnal flows, the residence
timescale is tightly connected to the daily cycle and could be of a
few hours during the night, or the entire night. The transition
between two equilibria typically takes place over a duration of about
a half hour. For reliable statistical estimation, multiple tests
showed that a minimum window length of 20 discrete points is
needed. With a sampling time of 1 minute, that means that a moving
window of approximately 20 or 30 minutes may be appropriate.

In addition, the sampling frequency has to be fine enough to sample
typical fluctuations of the dynamics. In the following analyses, we
find a sampling frequency of one minute to be appropriate for that
purpose. The characteristic timescale here is given by the timescale
at which the system recovers from perturbations (which is estimated by
linear stability analysis in the case where the model is known, see
e.g.  \citep{vandeWiel:2017ju}), and the time interval between two
observations should be close to or smaller than this quantity
so that (small-scale) local fluctuations can be
  resolved. Since the characteristic timescales of the system cannot
be known analytically in many situations, for example when analyzing
time series from atmospheric models or from field data, we suggest two
data-driven approaches to select a window length:
\begin{itemize}
\item In the first approach, the mean residence time around each
  metastable state as well as the mean transition time between the two
  states will be estimated based on a \emph{data clustering
    approach}. The observations will be clustered in the metastable
  regimes and an intermediate, transition regime. From the clustered
  data, the mean residence time in each cluster will be
  evaluated. This approach will provide an upper bound to select the
  window length.
\item The second approach is motivated by the fact that the indicator
  $\Upsilon$ is obtained through a statistical inference procedure
  through the definition of the BIC which involves fitting suitable
  ARMA processes to data.
  Specifically, a maximum likelihood approach is used, which assumes
  that subsequences are sampled from a normal distribution.  To assess
  the validity of this \emph{statistical approach}, a normality test
  will be implemented as a criterion to select a window length for
  which the normality assumption is justified and ARMA model
  estimation is reliable.
\end{itemize}
Both approaches are applicable when the data-generating model is
unknown. This is important in cases where data showing signs of
metastability are available, but an underlying model is unknown. A
summary of the full algorithmic procedure used to calculate the
statistical indicator is given in Appendix~\ref{sec:algorithm}.

\subsubsection{Clustering Approach: K-means}\label{subsubsec:KMeans}
In the first approach, we suggest to use the K-means algorithm
(\cite{Hartigan1979KMeans}, see pseudo code in
Appendix~\ref{sec:kmeans:details}) to select a window length for the
analysis. In the context of analyzing transitions in the temperature
inversion, the idea is to cluster the data into three different
clusters: data around each stable fixed point and data near the
unstable fixed point (in other words, data covering the transition
periods between two metastable states). By that, the goal is to
estimate the average time needed by the system to transition between
two metastable states. The mean residence time within each cluster is
calculated from the time series of cluster affiliation.  We choose
$\tau$ (recall that we consider it in number of discrete observations
and not in physical time) such that it is smaller than the minimal
mean time spent in one cluster, which should ensure that subsequences
remain mostly around one equilibrium. This value is denoted by
$\tau_{Kmeans}$. For the simulated data in the following, each
simulated time series will be assigned a window length $\tau_{Kmeans}$ by this
procedure. For the nocturnal dataset, we cluster the entire dataset
once and obtain a length $\tau_{Kmeans}$ of 22 points, corresponding to a
duration of 22 minutes. For the Polar dataset, only one continuous
time series during a Polar winter will be considered and assigned one
value of $\tau_{Kmeans}$, namely 10 points, corresponding to a duration of 100
minutes. This window length is insufficient to obtain reliable
statistical estimations.

Note that this clustering approach to determining a residence
timescale around an equilibrium is a crude approximation and suffers
from many caveats: a high density of data close to a given value of
the temperature inversion may not necessarily relate to the existence
of metastable equilibria, but could occur due to non-stationary
dynamics or complex nonlinear effects, for example. Nevertheless, we
use it as a first approach and future research may result in more
reliable approaches. In the following analyses, the K-means procedure
can be interpreted as providing an upper bound for selecting a window
length for the analysis and thus, combined with the following
criterion, will offer an applicability criterion for our method.

\subsubsection{Statistical Approach: Anderson-Darling Normality Test}\label{subsubsec:ADtest}
The K-means clustering approach described above estimates the system's
physical timescales, but the statistical properties of the process
should also be considered for reliable calculations. To fit ARMA
models reliably and to calculate the Bayesian information criterion
for ARMA model selection, we need the underlying process to follow a
normal distribution. Consequently, we suppose that the subsequences
are sampled from a normal distribution, at least for some window
length $\tau$. We then choose $\tau$ as the biggest window length such
that this normality assumption holds (more precisely, such that the
normality hypothesis cannot be rejected). This value is denoted by
$\tau_{AD}$. Specifically, the statistical test results in a p-value
for each subsequence, and we choose the window length such that the
median of the p-values of all subsequences is above a threshold for
which the null-hypothesis cannot be rejected. The normality test
applied here is the Anderson-Darling Test \citep{Anderson1952ADTest},
abbreviated AD test, as it is, for example, more stable than the
Kolmogorov--Smirnov test \citep{Stephens}. Further details details of
the AD test are summarized in
Appendix~\ref{sec:ADtest:details}. Similarly to the clustering
approach, the window length $\tau_{AD}$ for which normality cannot be
rejected is selected for each analyzed time series, and this value of
$\tau_{AD}$ is then used for all subsequences of the time series. Each
of the simulated dataset, the nocturnal dataset and the Polar dataset
will be assigned a single value of $\tau_{AD}$. The
  value of $\tau_{AD}$ for the nocturnal dataset is 19 discrete points, hence
  19 minutes with a sampling frequency of one minute. For the Polar
dataset, the value of $\tau_{AD}$ is 43 points corresponding to a
duration of 430 minutes.

\section{Stability analysis of simulated and observed time series}
In this section we quantify the reliability of the stability indicator
introduced in section
\ref{sec:AnalysesDynStab}.\ref{subseq:StatIndicator}. We start by
testing it on a controlled dataset generated by the simplified model
for near-surface temperature inversion (see section
\ref{sec:AnalysesDynStab}.\ref{subseq:ModelDescription}) and then
proceed by applying $\Upsilon$, the stability indicator, to
observational data. In the tests we use the auto.arima() function from
the "forecast" R package \citep{forecastRpackage}. The auto.arima()
function fits ARMA(p,q) models by calculating the maximum likelihood
estimators for a given model order (using the innovation algorithm
mentioned earlier). It calculates the corresponding BIC (using the
definition (\ref{Eq:BIC})) for all ARMA(p,q) models with
$p\leq p_{max}$ and $q\leq q_{max}$, where $p_{max}$ and $q_{max}$ are
thresholds set to 10 in our application, and then it chooses the ARMA
model with the minimal BIC value. This procedure is repeated for each
subsequence of data, using the moving window approach, and the minimal
BIC value leads to the optimal ARMA(p,q) model to represent the given
subsequence.

\subsection{Simulated time series}\label{subsec:SimTimeSeries}
To generate the simulated data, we use the conceptual randomized model
\eqref{eq:ODEvandeWiel:SDE}, which we recall here for the reader's
convenience:
\begin{equation*}
dx(t)=-\frac{dV\bigl(x(t)\bigr)}{dx}dt+\sigma dB(t)\;,\quad x(t_0)=x\;,\quad 0\leq t\leq T\;,
\end{equation*}
where $V(x)$ is the energy potential defined
in~\eqref{eq:PotVanDeWiel}. That is, the data-generating model reads
\begin{equation}
	dx= \begin{cases} 
	 \left( Q_i-\lambda x-Cx(1-x) \right) dt +\sigma dB, & x\leq 1\;,\\ 
	  \left(Q_i-\lambda x \right) dt +\sigma dB, &  x>1\;.
	\end{cases}\label{eq:SDESimplModel}
\end{equation}
The SDE model~\eqref{eq:SDESimplModel} is approximated path-wise
(i.e., for each realization of the driving Brownian path) using the
Euler-Maruyama scheme.

\begin{figure}[hbt]
	\centering
	\includegraphics[width=0.5\textwidth]{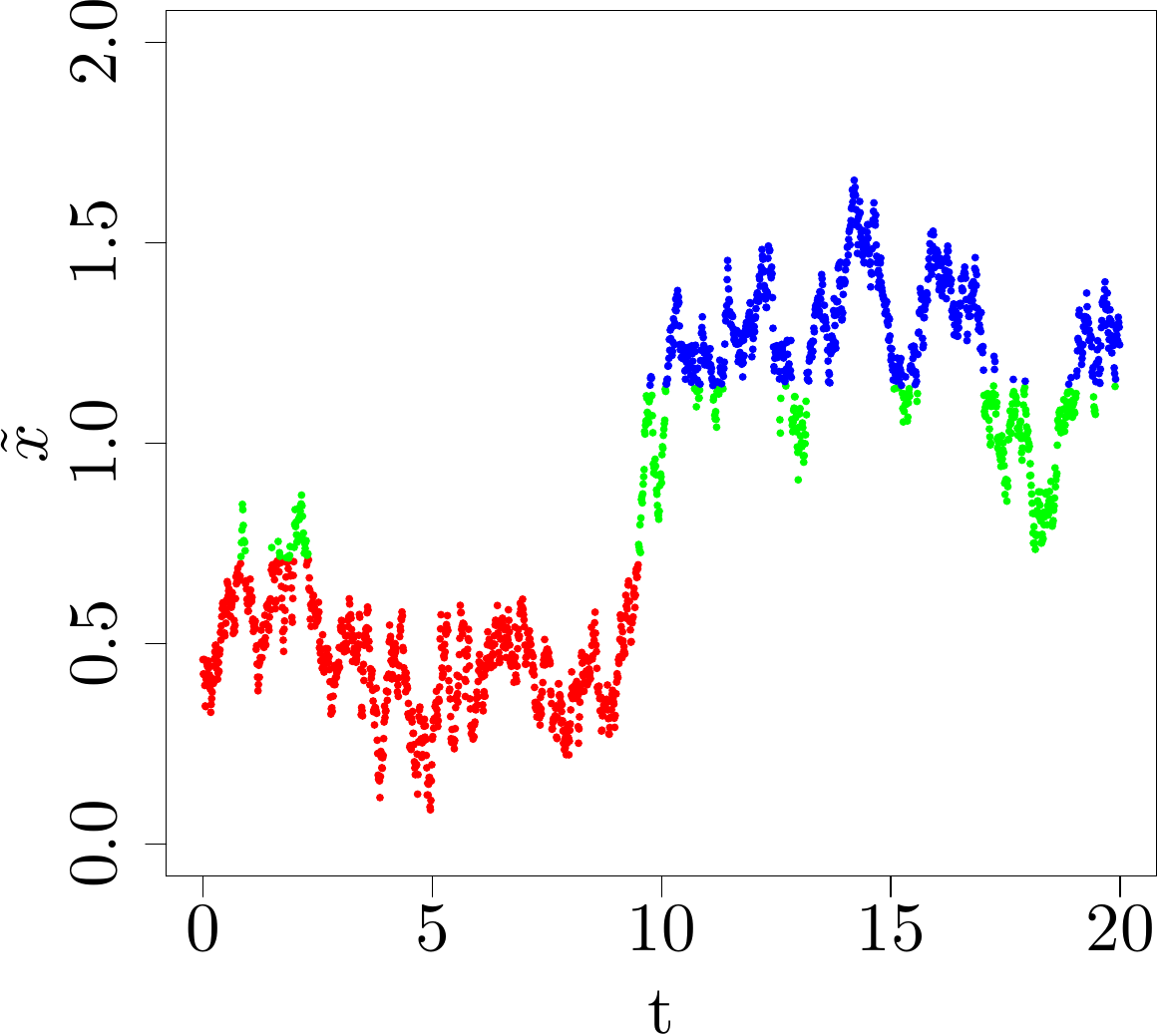}
	\caption{Clustered simulated time series for $C=6.4$ with K-means algorithm. All points with the same color correspond to the same cluster. The window length estimated by the K-means algorithm  is $\tau_{KMeans}=94$ timesteps, with a discrete timestep $\Delta t=0.01$. Hence the window duration corresponds to $0.94$ time units.} \label{fig:ClustContData}
\end{figure}

\begin{figure}[h]
	\centering
	\includegraphics[height=0.65\textheight]{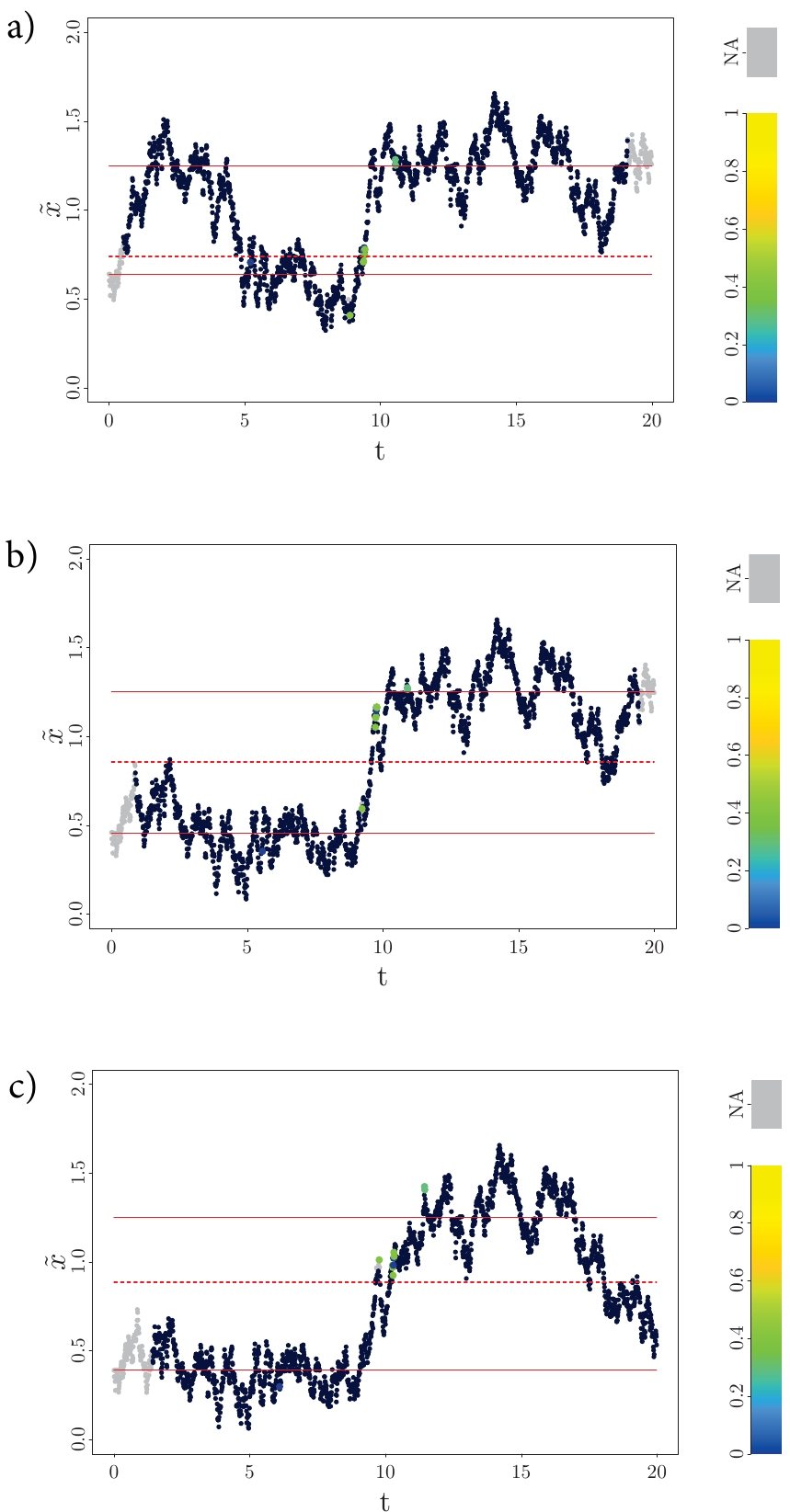}
	\caption{$\Upsilon$ for simplified model with a) $C=5.3$, b) $C=6.4$, c) $C=7.2$ and added noise, estimated for a window length a) $\tau = 54$, b) $\tau = 89$, c) $\tau = 143$ discrete timesteps. The grey dots correspond to missing values which are due to the fact that we only colour the last point of the modeled subsequence. Moreover, in rare cases the auto.arima() R function is not able to find an appropriate model. Red lines mark the stable equilibria, while the dotted red line marks the unstable equilibrium.}\label{fig:YContrData}
\end{figure}

For the purpose of testing the accuracy of the $\Upsilon$ indicator
and its potential to detect nearing regime transitions, one
realization $\{x_t\}$ of the stochastic process is used for each fixed
value of the bifurcation parameters $C$. Multiple fixed values of $C$
are used, resulting in one timeseries per value of $C$. The initial
parameters are set to $t_0=0$ and $x(t_0)=min\{x_{e_i} | i=1,2,3\}$
where $x_{e_i}$ are the three equilibria of the system. To generate
the controlled data set the model parameters are set to $\lambda= 2$,
$Q_i= 2.5$ and $\sigma = 0.35$. The value of $C$ is varied between
$C=5.3$ and $C=7.2$ with discrete increments of $0.1$ and one
simulation is done per value of $C$.  The simulations are ran for
$n = 2000$ time steps of size $\Delta t= 0.01$. The amplitude of the
noise, or diffusion coefficient, $\sigma=0.35$ is chosen as it
resulted in trajectories for which regime transition could be observed
on the time interval $[0,T=n \Delta t=20]$.  The range for $C$ is
chosen because for these values the time series shows frequent
transitions from one metastable state to another. To choose the window
length $\tau$ we apply both the K-means Algorithm (section
\ref{sec:AnalysesDynStab}.\ref{subseq:StatIndicator}.\ref{subsubsec:KMeans})
and the Anderson Darling Test (section
\ref{sec:AnalysesDynStab}.\ref{subseq:StatIndicator}.\ref{subsubsec:ADtest}). The
length $\tau$ is determined individually for each simulation, i.e. for
each fixed value of $C$. The K-means algorithm can be used to estimate
the amount of discrete observation points covering the transition
time. We set the cluster number to three as we expect three
equilibria. The results of the clustering algorithm are exemplary
shown in figure \ref{fig:ClustContData} for $C=6.4$. 
Note that
$t\in [0,n\cdot \Delta t]=[0,20]$, whereas we will express our window
length $\tau$ in number of discrete points in the following. In this
case the equilibria are
$x_{e_1}=0.46 \text{ (metastable)}, \, x_{e_2}= 0.97 \text{
  (unstable)}$, and $x_{e_3}=1.25 \text{ (metastable)}$. The cluster
centers, estimated by the K-means algorithm, are 0.46, 0.97, and 1.31
which are a close approximation of the equilibria. Therefore, we
expect a good estimation for the amount of points covering the
transition. The average time spend in each cluster are (for $C=6.4$):
$mean(T_1) = 112.2$, $mean(T_2)= 94$, and $mean(T_3)= 286.67$, where
$mean(T_i)$ is the average time spent without observed transitions in
cluster $i \in \{1,2,3\}$ expressed in number of discrete points. The
minimal mean residence time is thus $mean(T_2)= 94$ and provides an
upper bound to select a window length that respects the timescales of
the system. The window length $\tau$ is thus chosen such that it is
smaller than the minimal average time spent in one cluster, i.e. for
$C=6.4$ we choose
$\tau<\tau_{KMeans}:=\min\{mean(T_i)| i =1,2, 3\} = 94$. For all
tested $C$, we choose $\tau=\min\{mean(T_i)| i =1,2, 3\}-5$ in order
to give room for some uncertainty in the evaluation of the time spent
in each cluster, due to potential overlaps of the clusters (we recall
that the minimal mean residence time should be understood as an upper
bound to select $\tau$). By
  applying $\Upsilon$ to the data generated by the simplified model
  with $C=5.3$, $C=6.4$ and $C=7.2$ we get the results shown in figure
  \ref{fig:YContrData}. 
The solid red lines correspond to the stable
equilibria and the dotted red line to the unstable one. The colors
ranging from dark blue to yellow represent the stability of the points
measured by $\Upsilon$ and we always color the last point of the
subsequence. The simulation is initialized around the stable
equilibrium $x_{e_1}$, where short memory of the random perturbations
should prevail. As expected, the values of $\Upsilon$ remains close to
0 (corresponding to a most likely AR(1) model) as long as the
simulation oscillates around the equilibrium. The timeseries
eventually approaches the neighborhood of the unstable equilibrium
where long memory properties are to be expected and thus higher order
ARMA(p,q) models, hence larger values of $\Upsilon$, are more likely.
This first transition through the unstable equilibrium is well
recognized with higher values of $\Upsilon$ (green dots after the
dotted red line).
 
\begin{figure}[hbt]
	\centering
	\includegraphics[width=0.4 \textwidth]{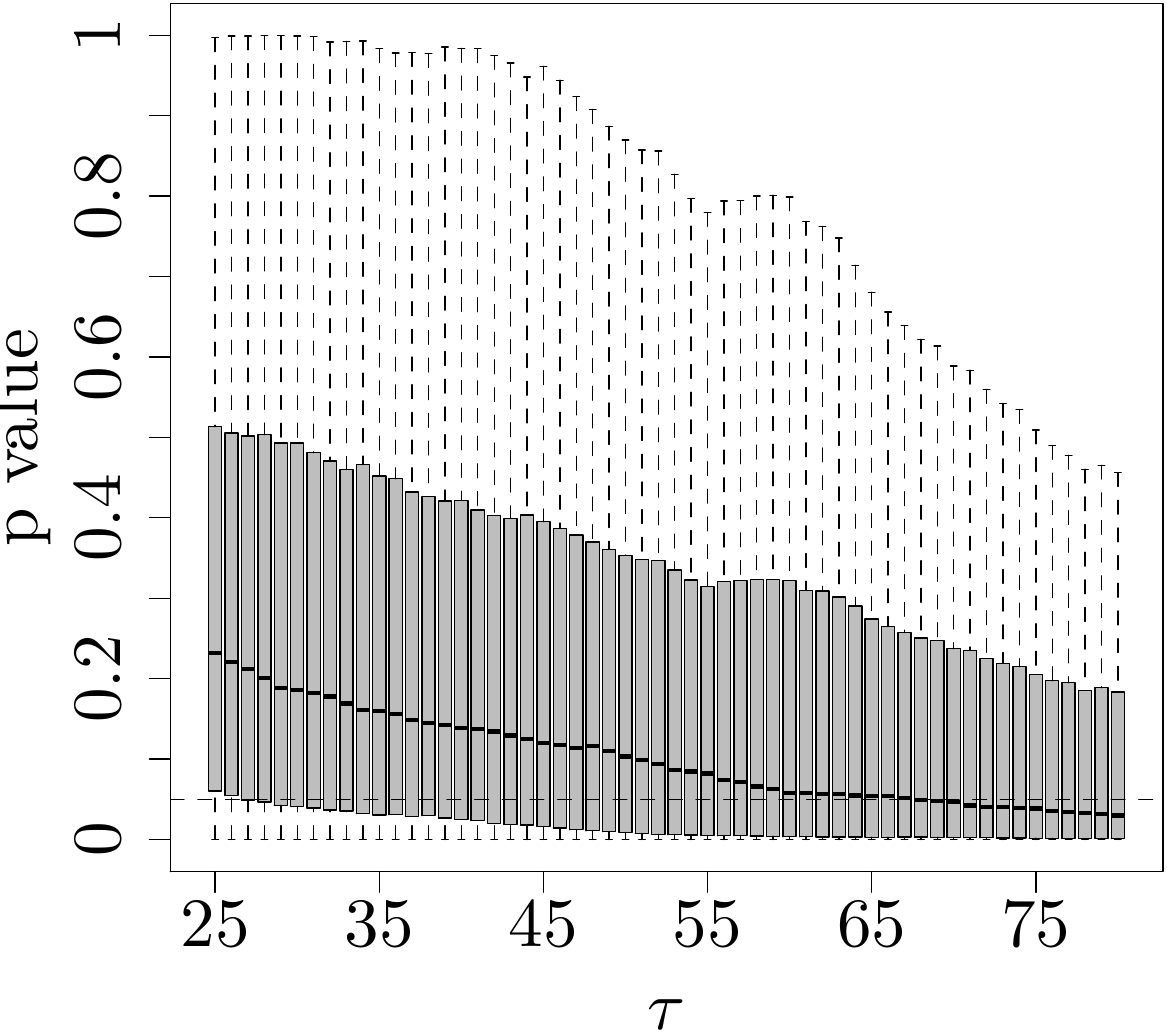}
	\caption{Boxplot of the p-values from the Anderson Darling Test for simulated time series with $C=6.4$. The window length estimated by the Anderson Darling test is $\tau_{AD}=67$ discrete timesteps.}\label{fig:BoxPlotContrData}
\end{figure}

The Anderson Darling Test can be used to find the biggest $\tau$ for
which we can assume that most of the subsequences are sampled from a
normal distribution and hence trust the ARMA model selection and
fitting. As shown in figure \ref{fig:BoxPlotContrData}, for $C=6.4$
the Anderson Darling Test yields that for $\tau=67$ the median of the
p-values for all subsequences is greater than the significance level
0.05. 
The solid line in the gray boxes is the median of the p-values
for a fixed $\tau$ while the upper and lower border of the gray boxes
refer to the upper and lower quartile of the p-values. The dotted
horizontal line is the significance level. We report that the values
of $\tau_{AD}$ given by the Anderson Darling Test are ranging from 60 to 70
discrete points for all values for $C$.

\begin{figure}[hbt]
	\centering
	\includegraphics[width=0.9\textwidth]{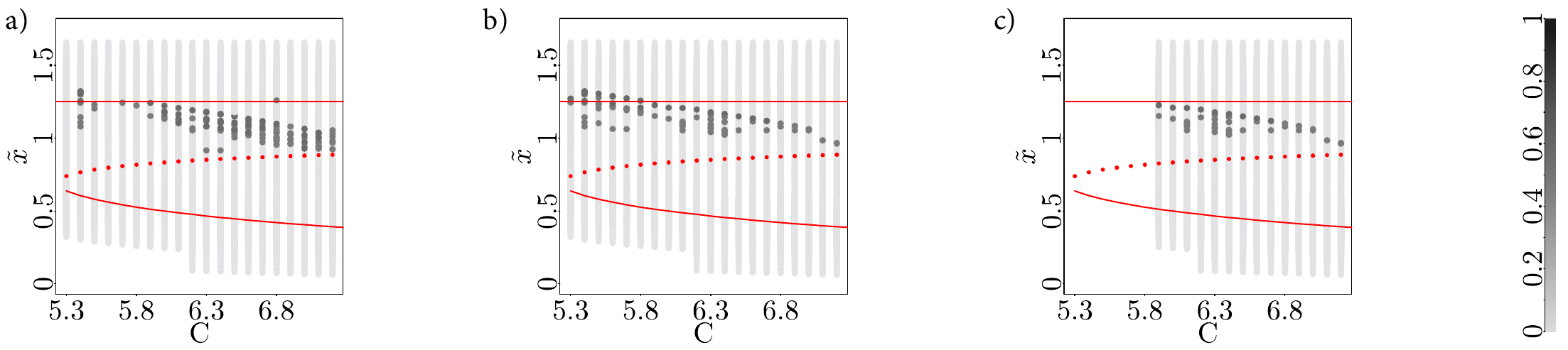}
	\caption{Bifurcation diagram of the deterministic system (red) and $\Upsilon$ calculated for simulated data (gray dots): a) for $\tau = \min\{mean(T_i)| i =1,2, 3\}-5$ (K-means), b) for $\tau=\tau_{AD}$ (Anderson Darling Test) and c) $\tau=\tau_{AD}$ if $\tau_{AD}<\tau_{KMeans}-5$, otherwise the subseries are discarded. The red full and dotted lines show respectively the stable and unstable branches of the bifurcation diagram. }\label{fig:BifPlotContrData}
\end{figure}

Figure \ref{fig:BifPlotContrData} summarizes the $\Upsilon$ values
obtained for different choices of $\tau$ and different values of $C$
in a bifurcation diagram. 
In the figure, the equilibrium solutions of
the deterministic equation \eqref{eq:ODEvandeWiel} are shown by a red
line for the considered range of values of $C$. This is the same
diagram as shown in Fig. \ref{fig:fig_bifurcation}, where the upper
and lower branches of the equilibrium solution correspond to the two
stable equilibria, while the middle one is the unstable equilibrium
separating the two basins of attraction of the stable equilibria. A
discontinuity in the solution is visible between the upper and middle
solution branches, which is due to the discontinuity introduced by the
cutoff form of the stability function. The $\Upsilon$ values obtained
for the simulations of the stochastic system
(eq. \eqref{eq:SDESimplModel}) for all considered values of $C$ are
then shown as a scatter plot along with the equilibrium solution, and
the darker color corresponds to higher values of $\Upsilon$. As the
initial condition for all simulations is taken at the lowest
equilibrium values, the transitions are expected to occur between the
lower and upper equilibrium branches when the system transitions from
the basin of attraction of the lowest equilibrium value to that of the
highest value. High values of $\Upsilon$ are indeed mainly found in this region of the diagram. 

Three methods are used to select the value of $\tau$ and the results
associated with these window sizes are shown in figure
\ref{fig:BifPlotContrData}. In figure \ref{fig:BifPlotContrData} panel
a, $\tau =\tau_{KMeans}-5$ is used. Around the stable branches, values
of $\Upsilon$ are small, denoting that stable states are detected as
such. Large values of $\Upsilon$ are found between the unstable branch
and the upper stable branch of the bifurcation diagram, indicating
that transitions from the lower to the upper stable branches are
detected by the indicator. The fact that the high values are not
exactly located around the unstable branch is due to the use of a
finite window size for the calculation: in the diagram, the color is
always assigned to the last point of the subsequence. For small values
of $C$, e.g. for $C=5.4$, large values of $\Upsilon$ are occasionally
inappropriately found around the upper stable branch. For small $C$,
the potential well will be relatively steep and the system rapidly
approaches the second equilibrium, so that the detection can be too
slow. Figure \ref{fig:BifPlotContrData} panel b shows the results for
$\Upsilon$ when choosing $\tau$ according to the Anderson Darling
test, denoted as $\tau_{AD}$. The figure is very similar to the one
using $\tau_{KMeans}$ except that for $C\leq 5.6$ there are more high
values of $\Upsilon$ located around the stable branch. This is due to
the fact that for these $C$'s the $\tau$'s chosen by the Anderson
Darling test are larger than the ones estimated by the K-means
algorithm. Consequently, the local stationarity assumption may break
down. For $C\geq 5.9$ the $\tau$'s given by the AD Test are smaller
than the ones of the K-means algorithm. In these cases $\Upsilon$
gives a good indication for the stability. Figure
\ref{fig:BifPlotContrData} panel c is a bifurcation plot showing only
the time series for which $\tau$ can be chosen to satisfy both the
K-means and the Anderson Darling condition,
i.e. $\tau_{AD}<\tau_{KMeans}-5$. Here $\tau:=\tau_{AD}$ is used for
the analysis and timeseries that do not satisfy the condition are
discarded from the analysis. In this case, we see that large values of
$\Upsilon$ always occur between the unstable branch and the upper
stable branch, thus $\Upsilon$ is capable of recognizing the location
of unstable equilibria for all $C$ and stable equilibria are never
assigned a large value of $\Upsilon$. Therefore, we are confident that
we can apply $\Upsilon$ to observational data. Moreover, for this
simple bistable example system, analytical results can provide the
expected time taken by the system to transition from one of the local
equilibria to the bifurcation point \citep{MFPTPaper} and can serve as
a comparison to the statistical estimations of $\tau$ obtained
here. As a matter of fact, for $C$ close to the bifurcation point, the
results given by the Anderson Darling test are similar to those given
by analytical calculations.

\subsection{Analysis of regime transition in observed nocturnal and Polar temperature inversions}
\begin{figure}[hbt]
	\centering
	\includegraphics[width=0.9\textwidth]{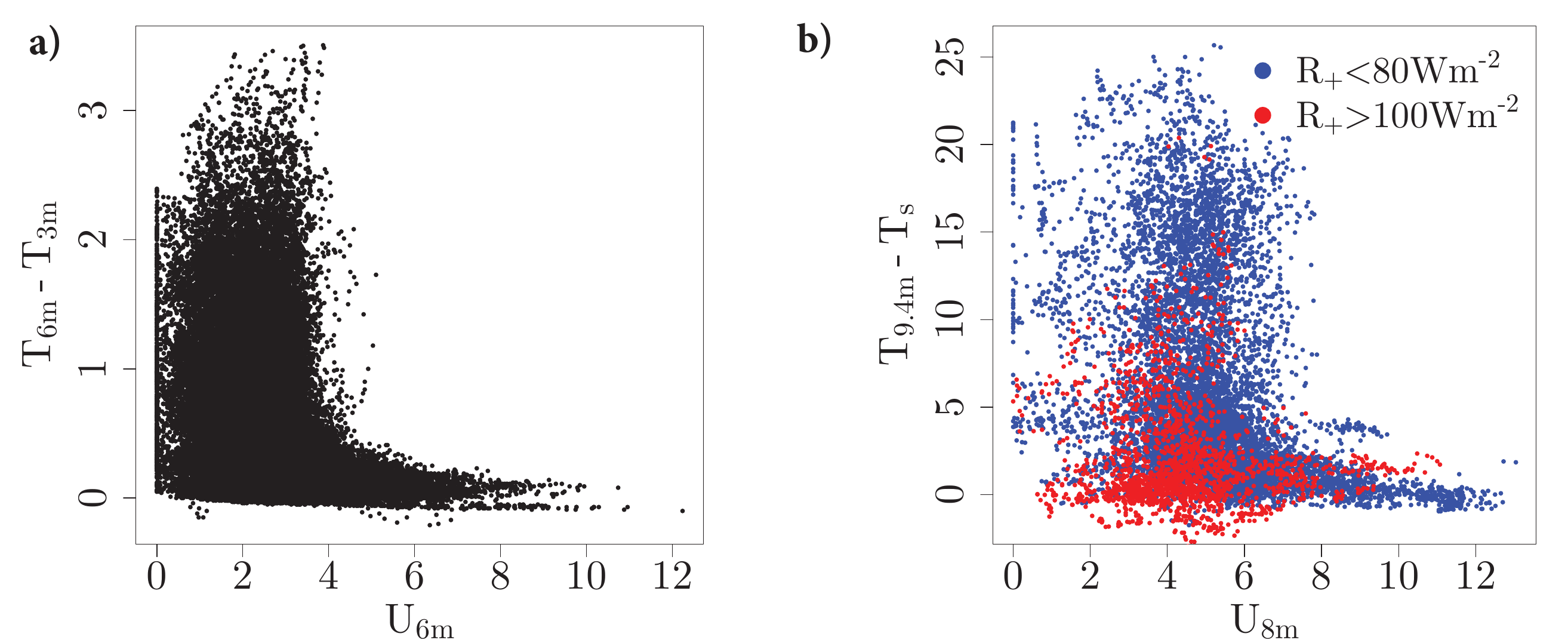}
	\caption{Temperature inversion as a function of wind speed as observed at Dumosa, Australia (a) and Dome C, Antarctica (b). The color in panel b corresponds to lower and higher incoming radiation.}\label{fig:TempInvDumDomeC}
\end{figure}

In this section we apply the stability indicator $\Upsilon$ on
observational data obtained from one site near Dumosa, Australia for
which nocturnal data are selected, and from Dome C, Antarctica for
which we consider the Polar winter. When we plot $\Delta T$ over $U$
for both sites (see figure \ref{fig:TempInvDumDomeC}) we see a clear
sign of two metastable states: one when the wind is weak and
$\Delta T$ is large and one for strong wind where $\Delta T$ is small.

\subsubsection{Dumosa}
\begin{figure}[hbt]
	\centering
	\includegraphics[width=0.9\textwidth]{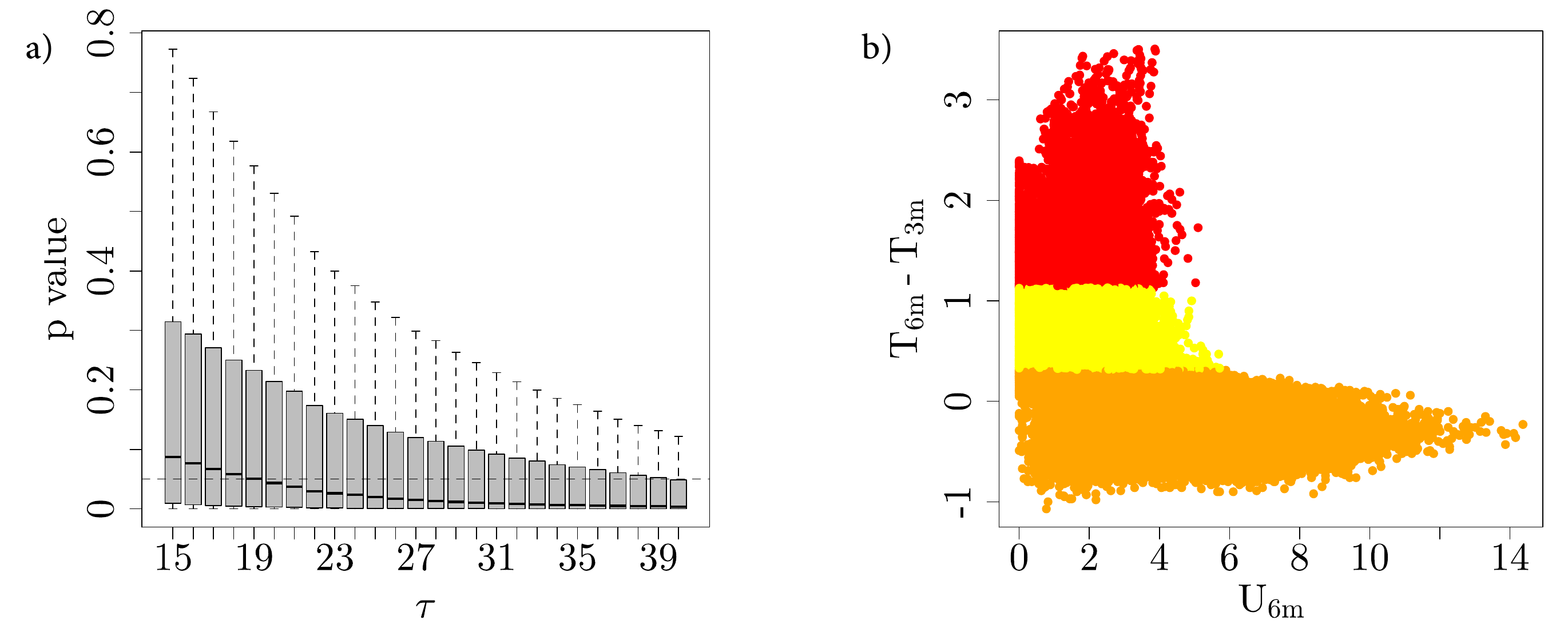}
	\caption{a) Boxplot of the p-values from the Anderson Darling test and b) Clustered data with K-means for the Dumosa data, for all 89 nights.} \label{fig:boxPlotClusterDumosa}
\end{figure}

\begin{figure}[hbt]
	\centering
	\includegraphics[width=0.9\textwidth]{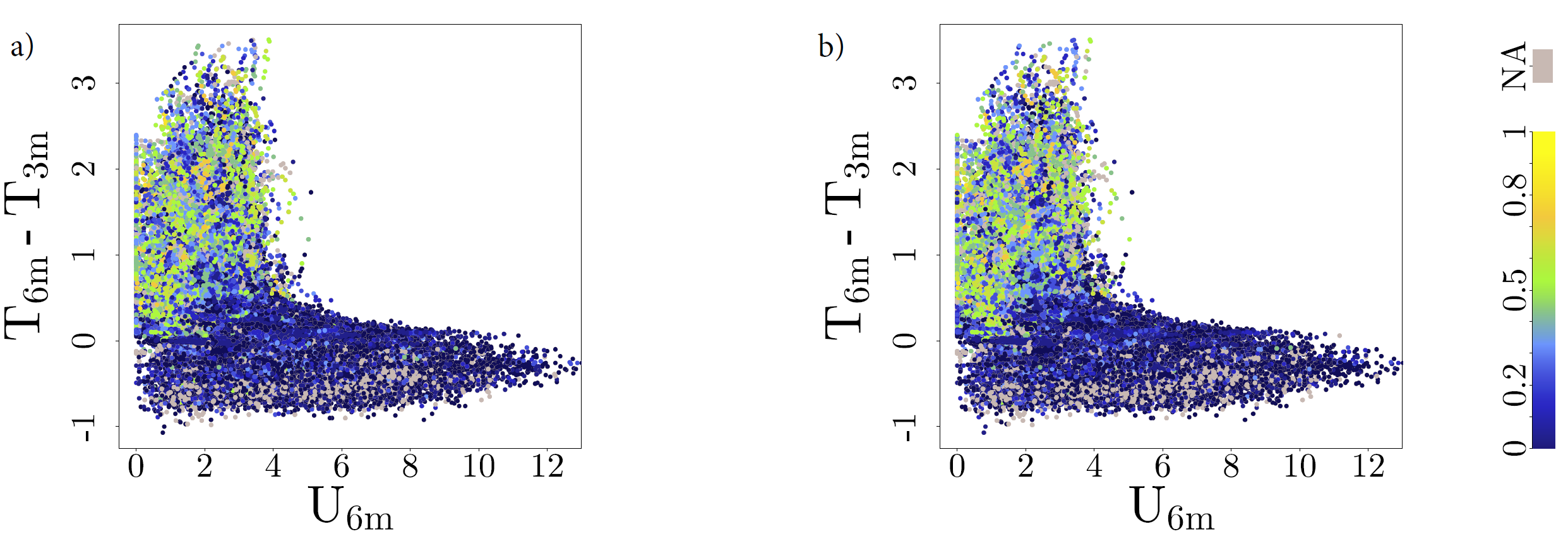}
	\caption{Observed temperature inversion versus wind speed relation for the Dumosa data. Colored according to $\Upsilon$ with different window lengths $\tau$: a) $\tau= \tau_{AD}$, b) $\tau = \tau_{KMeans}$. The grey dots correspond to missing values which are due to the fact that we only color the last point of the modeled subsequence. Moreover, in some cases the auto.arima() R function is not able find an appropriate model.} \label{fig:BifurcationDumosa}
\end{figure}

\begin{figure}[hbt]
	\centering
	\includegraphics[width=0.9\textwidth]{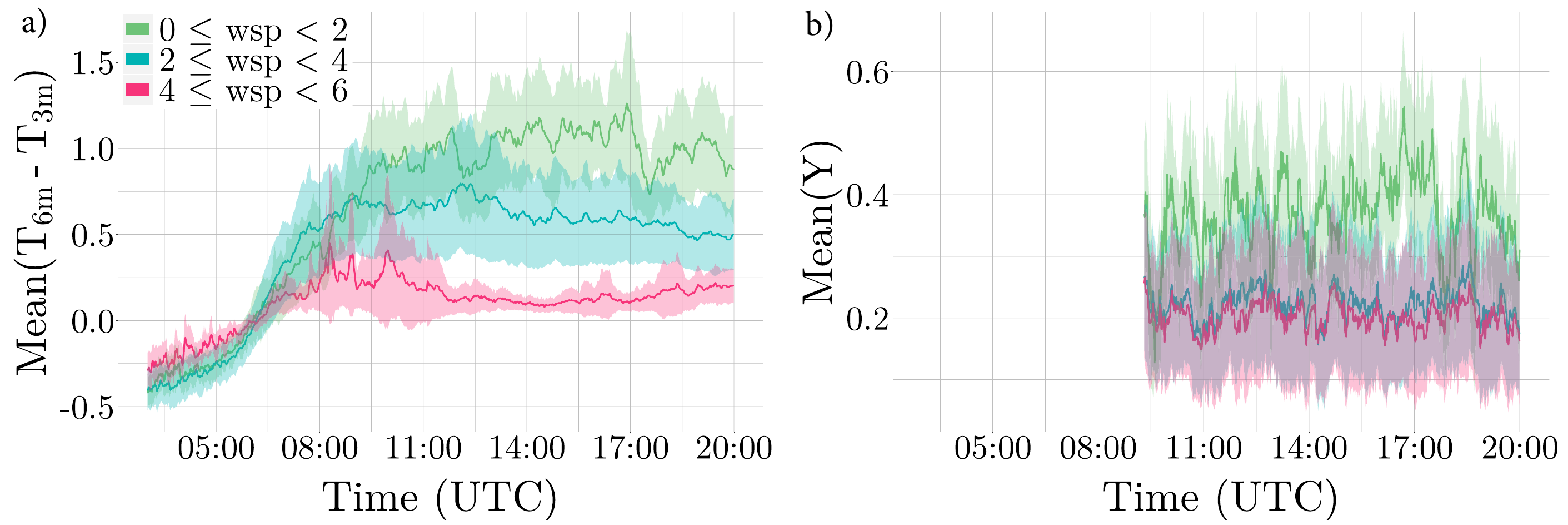}
	\caption{Time series of mean($\Delta T$) (a) and mean($\Upsilon$) (b) for different wind speed (wsp) categories for Dumosa data, calculated with $\tau_{KMeans}$. The shaded area is the standard deviation. } \label{fig:MeanWindDumosa}
\end{figure}

The first observational dataset consists of temperature measurements
from a site near Dumosa, Victoria, Australia. The site was located in
a large area with mostly homogeneous and flat terrain, covered by
wheat crops, and measurements were taken during the crop season. The
temperature measurements were made on the main tower at heights of 3
and 6m and the wind measurements at 6m. The frequency of measurements
is 1 minute. Further details about the observational site can be found
in \cite{Lang:2017bua}. As we want to use data where we can expect
temperature inversions to take place we exclusively use evening and
nighttime data from March until June 2013 (89 days). Each night of
data results in a timeseries of 1020 discrete observations. Similarly
to the simulated data, we use the K-means algorithm and the Anderson
Darling Test to choose the window length $\tau$. The
  results are shown for all nights considered together in Figure
  \ref{fig:boxPlotClusterDumosa}. 
According to these tests the maximal
  $\tau$ for which we can assume normality ($\tau_{AD}$) is 19
  discrete observations (hence 19 minutes with the sampling frequency
  of 1 minute) and the $\tau$ which corresponds to the minimal mean
  residence time in one of three clusters ($\tau_{KMeans}$) is
  22. Hence we have $\tau_{AD} < \tau_{KMeans}$ and choosing
  $\tau_{AD}$ should be appropriate. The results for $\Upsilon$
  applied to all 89 nights with both choices of window lengths are
  given in figure \ref{fig:BifurcationDumosa}. 
The results highlight a
  lower branch with low values of $\Upsilon$, or dynamics identified
  as stable, and an upper branch with high values of $\Upsilon$, or
  dynamics identified as unstable. In some cases, a proper ARMA(p,q)
  model cannot be fitted by the statistical methods, resulting in absence of results for some
  windows. Generally, a reliable ARMA(p,q) fit becomes difficult for a
  time series with less than 20 observations, and the estimated window lengths are on the lowest end to
  obtain the statistical estimations. Figure \ref{fig:MeanWindDumosa}
  shows the time evolution of $\Delta T$ when conditionally averaged
  for all nights with the wind speed (wsp) being in a given
  category. The corresponding time evolution of $\Upsilon$ is shown
  for the same conditional averages. The window length here is chosen
  as the most restrictive criterion
  $\tau=\tau_{AD}<\tau_{KMeans}$. 
For low wind speeds, the $\Upsilon$
  values are high (on average), which implies that in this case we
  have an unstable system. Note that in this dataset, a leveling-off
of the temperature inversion for low wind speeds (which could
correspond to the stable equilibrium of a strong inversion according
to the model of \cite{vandeWiel:2017ju}) is not very evident from
Figure \ref{fig:BifurcationDumosa}. It could be that the temperature
inversion does not have time to reach the stable equilibrium during
the night, or that other instabilities which are not considered in the
simplified model arise in strong stability conditions. For example,
flow instabilities such as submeso motions, which are favored in
strongly stratified situations, could make the system dynamically
unstable.

\subsubsection{Antarctica}
\begin{figure}[hbt]
	\centering
	\includegraphics[width=0.9\textwidth]{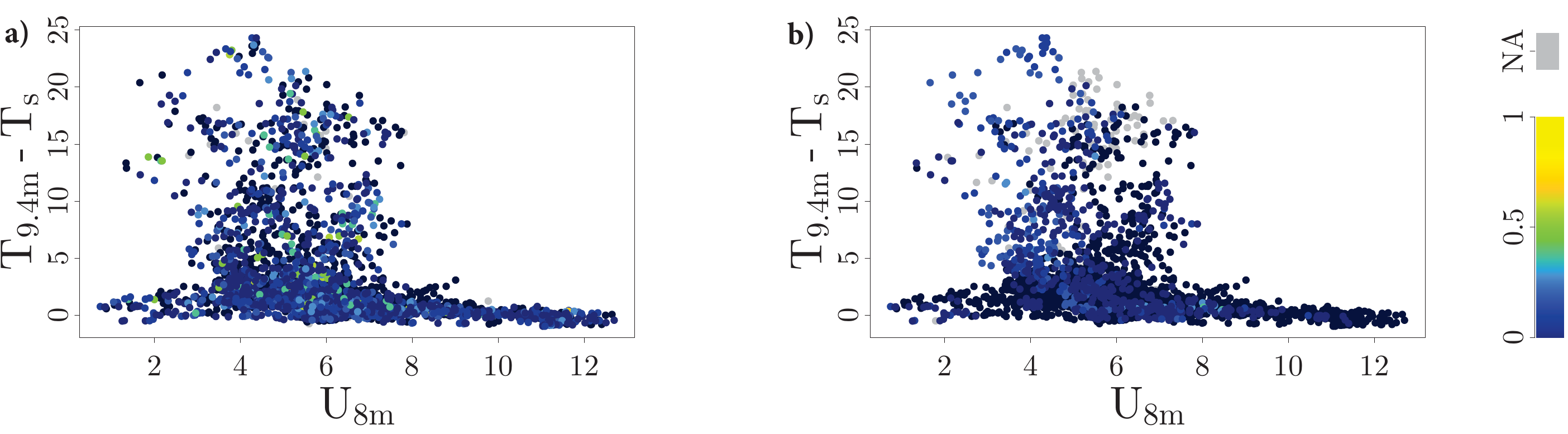}
	\caption{Temperature inversion between 9.4 m and the surface, as a function of wind speed at 8m as observed at Dome C. Colored according to $\Upsilon$ with different window lengths $\tau$, expressed in number of discrete observations: a) $\tau = 10$ (K-means) and b) $\tau=43$ (AD Test)} \label{fig:BifurcationDomeC}
\end{figure}

The Dome C data was measured at the Concordia Research Station which
is located on the Antarctica Plateau. It is a French-Italian research
facility that was built 3233m above sea level. It is extensively
described for example in \citet{Genthon2010DomeC}. The Dome C dataset
contains 10-min averaged meteorological data from 2017. Regimes and
their transitions were analyzed by \cite{Vignon:2017bf} and
\cite{Baas:2019cl}.  Important for our analysis are measurements of
the temperature at height 9.4 m and surface, the wind speed (m/s) at
height 8 m and the radiation made in the polar night which is from
March to September. We focus on the polar night during which multiple
regime transitions take place. Following \cite{vandeWiel:2017ju} the
data is classified into two subcategories of radiative forcing being
the sum of net shortwave and incoming longwave radiation:
$R_+=K^{\downarrow}-K^{\uparrow}+L^{\downarrow}$. Strong cooling is
favored in cases of low incoming radiation and when plotting
$\Delta T = T_{9.4m}-T_s$ over the wind speed $U_{8m}$ a back-folding
of the points becomes apparent when $R^{+}<80 Wm^{-2}$
(\cite{vandeWiel:2017ju}, their Figure 6 and less clearly in our
Figure \ref{fig:TempInvDumDomeC}). Therefore, we focus on the case
when $R^{+}<80 Wm^{-2}$. We apply $\Upsilon$ to the longest
consecutive time series with $R^{+}<80 Wm^{-2}$ which is from
2017-08-03 10:50 to 2017-08-24 21:50, i.e. 3091 data points. Our
analyzed time series is thus shorter than the one visualized in
\cite{vandeWiel:2017ju}, which explains the differences in the
scatter plot. Again we choose $\tau$ with the Anderson Darling Test and
the K-Means algorithm. The value for $\tau$ given by the Anderson
Darling Test $\tau_{AD}=43$ observations (with a corresponding window duration of 430 minutes) is much larger than the one given by the
K-means algorithm ($\tau_{KMeans}=10$ observations, corresponding to a window duration of 100 minutes), which is in fact too few points to
expect a good fit for the ARMA(p,q) model. In figure
\ref{fig:BifurcationDomeC} we see that no transitions are recognized
by $\Upsilon$ with $\tau=\tau_{AD}$ (left) but with
$\tau=\tau_{KMeans}$ (right) some transitions are noted. 
This
indicates that the data frequency of this data set is not high enough
to give reliable results for the stability indicator. 
\subsection{Sensitivity analysis on averaged data}
\begin{figure}[hbt]
	\centering
	\includegraphics[width=0.9\textwidth]{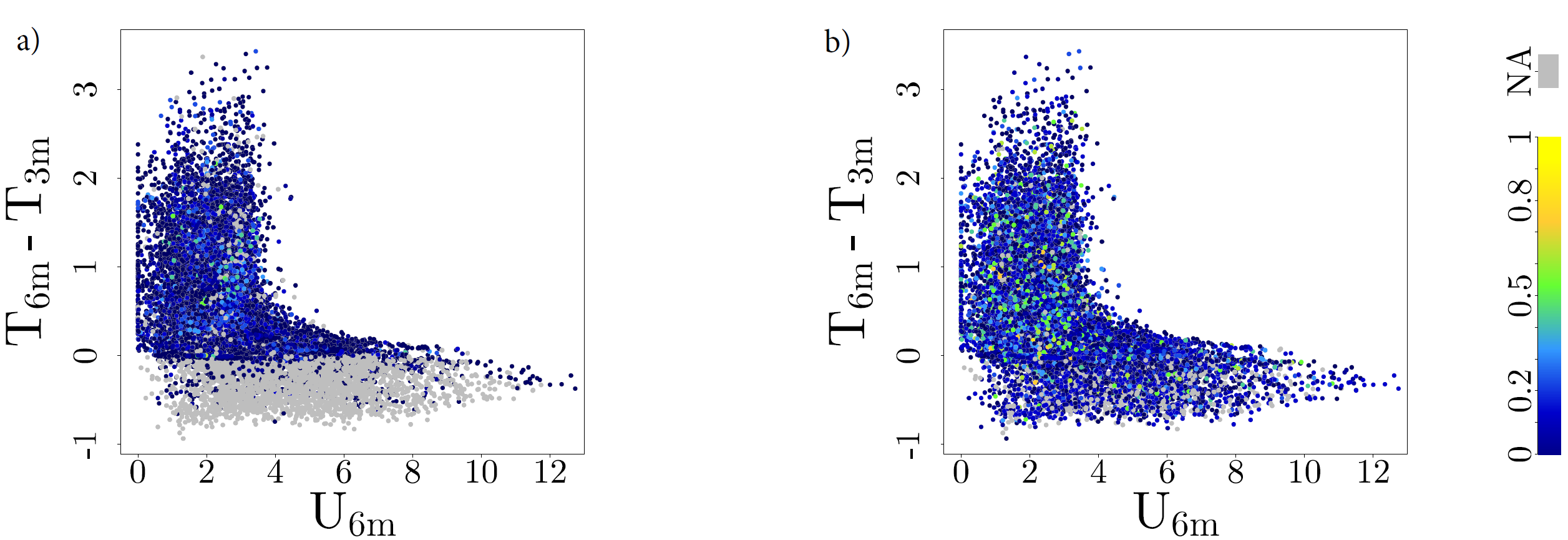}
	\caption{Observed temperature inversion versus wind speed relation for the 5-min averaged Dumosa data. Colored according to $\Upsilon$ with different window lengths $\tau$, expressed in number of discrete observations: a) $\tau= \tau_{AD}$ and b) $\tau = \tau_{KMeans}$.} \label{fig:BifurcationDumosa5min}
\end{figure}

As discussed when applying the indicator $\Upsilon$ on the Dome C
dataset, there is strong indication for the data frequency being
crucial for the reliability of the results when applying the stability
indicator. Observational data is often stored in block averages,
e.g. measurements over 5 minute time window are averaged into 1 data
point. The issue with this can be that the data frequency can be too
low to sample typical fluctuations during the observed transition. In
more detail, if the time taken by the system to transition from one
metastable state to the next is less than approximately 20 discrete measurement
points (the minimum needed to have relevant statistical results
according to our tests), then the approach may not be
applicable. Therefore, data frequency needs to be high enough to give
reliable results for $\Upsilon$. As a comparative study to illustrate
this point, we block average the temperature measurements for the
Dumosa data, such that we repeat the analysis based on 5-min averaged
data instead of 1-min averages. Thereby, we reduce the length of the
time series for each individual night from 1020 to 204 data
points. Again we choose $\tau$ with the Anderson Darling Test and the
K-means algorithm. There is a clear distinction between the $\tau$
estimated by the Anderson Darling Test and the one given by the
K-means algorithm for the 5 min data, contrary to the 1 min data where
both methods suggested comparable window lengths. Indeed, $\tau_{AD}=31$ and
$\tau_{KMeans}=7$ for the 5 min averaged data whereas $\tau_{AD}=19$
and $\tau_{KMeans}=22$ for the 1 min data. The small value for $\tau$
given by the K-means algorithm for the 5 min averaged data suggests
that there is only a small amount of points covering the transition
time and we cannot fit an ARMA(p,q) model properly to subsequences
this short. Moreover, as the value for $\tau$ given by the Anderson
Darling Test is much bigger than the amount of points covering the
transition time we do not expect reliable results for $\Upsilon$ with
this $\tau$. Figure \ref{fig:BifurcationDumosa5min} confirms this
hypothesis. The left plot is with $\tau=\tau_{AD}$ and the right one
with $\tau=\tau_{KMeans}$. 

\section{Discussion and conclusion}
In this study we analyzed the potential of a statistical indicator to
be used to detect the system's proximity to critical
regime transitions in the near-surface temperature inversion. The
statistical indicator evaluates the dynamical stability of time series
resulting from a dynamical system and was initially suggested in
\cite{Faranda_2014}. Based on idealized numerical simulations,
\cite{vanHooijdonk:2016bw} had found the presence of early-warning 
signs in the turbulent flow field before a transition from weakly
stable to strongly stable conditions. These signs included a critical
slowing down, referring to the fact that dynamical systems tend to
recover slower from perturbations when approaching a transition point
in the dynamics. This slowdown was evaluated based on fluctuations of
the temperature field and the early-warning signal relied on a change
in the variance. Such metrics, which are often used in studies of
tipping points, can become problematic when the underlying dynamics is
highly non-stationary, as an increase of variance could be due to the
non-stationarity of the system without implying a transition
\citep{lenton2012early,Faranda_2014}. The typical scatter of
atmospheric field data and their inherent non-stationarity makes the
application of classical critical slowdown metrics difficult. The
metric presented and used here is different in that it statistically
quantifies the deviation from the dynamics expected when the system is
close to a stable equilibrium. Specifically, the indicator is based on
ARMA modeling with a moving window for which local stationarity is
assumed, and the distance from stable equilibrium dynamics is
evaluated based on a Bayesian information criterion. The indicator
crucially relies on an appropriate window length and we suggested two
methods to select its value in a data-driven manner. That is, both
methods can be used when the underlying model governing the dynamics
is unknown, such that those can be applied to field data with
significant scatter. Our suggestion to ensure reliable results is to
use a combination of both approaches. The shortest residence time
around an equilibrium estimated through the K-means approach provides
an upper bound to select a window length that respects the timescale
of the system, i.e. a length that ensures local stationarity for ARMA
model fitting. The window length should be selected as shorter or
equal to this upper bound, and such that the data within individual
windows mostly satisfy normality to ensure reliable Bayesian
inference. The Anderson Darling normality test is appropriate, but an
improvement of the clustering approach to estimate the residence time
around an equilibrium (here done based on a simple K-means clustering
approach) would be beneficial. Based on this approach, we find that a
nocturnal temperature inversion dataset with a sampling frequency of 1
minute can be analyzed successfully using a window length of
approximately 20 minutes. Slower sampling frequency did not lead to
conclusive results.

The conceptual model introduced by \cite{vandeWiel:2017ju} was
developed to understand regime transitions in the near-surface
temperature inversion and can support scenarios with multiple stable
equilibria. For our purpose of identifying the system's
  proximity to  
regime transitions, it offers an ideal model for which the theoretical
dynamical stability can be calculated analytically. We extended the
model to include random perturbations in the dynamics and used the
resulting stochastic model to provide a test dataset on which to
evaluate the potential of the indicator of regime transitions. In this
stochastic system, small-scale perturbations can be amplified due to
the nonlinearity, resulting in transitions between the bi-stable
equilibria. Our simulations show such noise-induced regime
transitions, successfully identified by the indicator $\Upsilon$. More
research would however be beneficial in order to assess the type of
noise that is appropriate to represent randomized dynamics of the SBL.

The application to field data was done for one nocturnal dataset and
one Polar dataset. In their discussion, \cite{vandeWiel:2017ju} suggest
that the strength of the thermal coupling between the soil and the
atmosphere may be a key process to distinguish between cases where the
temperature inversion has a unique stable equilibria and cases with
bi-stable equilibria, separated by an unstable equilibrium. The
wind-speed dependence of observational scatter is partly attributed to
the existence of a dynamically unstable branch in the system in cases
where the thermal coupling is weak. In both datasets considered in our
analysis, a weak thermal coupling is to be expected. Clearly in the
Polar dataset, the snow surface leads to a weak thermal coupling
between the atmosphere and the soil \citep{Vignon:2017bf,
  vandeWiel:2017ju}. The nocturnal dataset originates from a wheat crop near Dumosa, Australia, probably resulting in a
weak thermal coupling as well. While the Dome C data did not have the
required sampling rate in order to have reliable estimates of the
dynamical stability, the Dumosa data were found to have a clear signal
with one dynamically stable branch and one dynamically unstable
branch. A second dynamically stable branch corresponding to a strong
inversion was not clearly observed. This data-driven result agrees
with the theoretical result of \cite{vandeWiel:2017ju}, namely that a
dynamically unstable branch exists for a certain range of wind speeds
in case of weak atmosphere-surface thermal coupling. Note that this is an idealized model and other non-represented physical processes may be at work and impact the interpretations.
This result is
nevertheless promising for the use of the indicator as an early-warning signal
of regime transitions. Extending the analysis to a Polar night with an
appropriate sampling frequency would be very interesting, as multiple
regime transitions occur during the long-lived temperature inversion
\citep{Baas:2019cl}. Moreover, comparing results obtained for a site
with strong atmosphere-surface thermal coupling would provide great
insight to compare the dynamical stability of field data to the
dynamical stability predicted by the conceptual model.

To be noted is the fact that the conceptual model is derived for a temperature inversion between the surface and a height at which the wind speed stays relatively constant during the night, found to be approximately 40 m at Cabauw in the Netherlands and 10 m at Dome C. The Dumosa dataset did not offer the possibility to select a height with such a constant wind speed. The measurements, taken at lower heights in this case, will be prone to submeso-scale activity, inducing perturbations of the shallow inversion which could affect the dynamical stability of the time series. In fact, the earlier application of the dynamical stability indicator to SBL data in \cite{Nevo2017StabInd} showed that higher stability corresponded to unstable dynamics of the vertical velocity fluctuations and of the wind speed. More analyses would be needed to assess the influence of the measurement height on the evaluated dynamical stability of the temperature inversion. Nevertheless, our results encourage the use of the statistical dynamical stability as a metric to detect nearing regime transitions in the SBL. The ability to detect nearing regime transitions in atmospheric numerical weather prediction and climate models could offer a possibility to use a different type of SBL parameterization in those specific cases without relying on the assumption of turbulence stationarity.


%
\section*{Acknowledgements}
The authors wish to thank Christophe Genthon and Etienne Vignon for providing the Dome C data, obtained as part of the CALVA observation project supported by the French polar institute IPEV, and for lively discussions. The research was funded by the Deutsche Forschungsgemeinschaft (DFG) through grant number VE 933/2-2. A.K acknowledges funding from the DFG through the Collaborative Research Center CRC1114, "Scaling Cascades in Complex system". The scientific exchanges between N.V and D.F were greatly facilitated by funding through the DAAD exchange program Procope 
through the project "Data-driven dynamical stability of stably stratified turbulent flows". We are grateful to anonymous reviewers for insightful comments that helped us improve the manuscript.

%
\section*{Appendix}

\subsection{Details of the K-means clustering algorithm} \label{sec:kmeans:details}

The clustering is done using the K-means algorithm with the following steps:

\begin{itemize}
	\item Input: k = \# number of clusters, set of points $x_{i-\tau+1}, \dots, x_{i}$
	\item Place centroids $c_1, \dots, c_k$ at random locations.
	\item Repeat until none of the cluster assignments change: 
	\begin{itemize}
		\item for each point $x_i$ find nearest centroid $c_j$ and assign $x_i$ to cluster $j$
		\item for each cluster $j=1, \dots, k$ calculate new centroid $c_j =$ mean of all points $x_i$ assigned to cluster $j$ in previous step.
	\end{itemize}
\end{itemize}

\subsection{Details of the Anderson-Darling Normality Test} \label{sec:ADtest:details}

The the Anderson-Darling (AD) normality test statistic is based on the
squared difference between the empirical distribution function
estimated based on the sample, $F_n(x)$, and the normal distribution
$F^{*}(x)$. The statistic for this test is,
\[W_n^2=n\int_{-\infty}^{\infty}[F_n(x)-F^*(x)]^2\psi(F^*(x))dF^*(x)\]
where $\psi$ is a non-negative weight function which is used to emphasize the tails of the presumed distribution. We use the modified AD
statistic given by D'Agostino and Stephens (1986) which takes into
accounts the sample size $n$ \[W_n^{2*}=W_n^2(1+0.75/n+2.25/n^2).\]
The null hypothesis of this normality test is that the data are
sampled from a normal distribution.  When the p-value is greater than
the predetermined critical value ($\alpha=0.05$), the null hypothesis
is not rejected and thus we conclude that the data is normally
distributed. 

\subsection{Summary of the algorithmic procedure} \label{sec:algorithm}

The full procedure to apply the statistical indicator to a timeseries follows the following steps:

\begin{itemize}
	\item Input: Timeseries $x_{1}, \dots, x_{T}$
	\item Evaluate the window length $\tau_{Kmeans}$:
	\begin{itemize}
		\item Cluster the timeseries in $k$ clusters (expected number of equilibrium states) using the K-means algorithm.
		\item For each cluster, calculate the mean residence time of the timeseries.
		\item The minimal mean residence time over the $k$ clusters provides an upper bound for $\tau_{Kmeans}$.
	\end{itemize}
	\item Evaluate the window length $\tau_{AD}$:
	\begin{itemize}
		\item For a range of window lengths $\tau_{min}<\tau<\tau_{max}$, apply the AD test statistic to the timeseries in a moving window approach. 
		\item For each $\tau$, calculate the median of the p-values obtained over all windows. 
		\item Select the largest $\tau$ for which the median p-value is greater than the predetermined critical value ($\alpha=0.05$) as $\tau_{AD}$.
	\end{itemize}
	\item If $\tau_{AD} < \tau_{Kmeans}$, select $\tau_{AD}$ as a window length. Else, ARMA model fitting may be innappropriate.
	\item Repeat for each window of length $\tau_{AD}$:
	\begin{itemize}
		\item Select the best fitting ARMA(p,q) model (minimal BIC).
		\item Fit an ARMA(1,0) to the window and calculate the BIC.
		\item Calculate $\Upsilon$. High values will indicate transitions. The values themselves may depend on the dataset.
	\end{itemize}
\end{itemize}

\bibliographystyle{plainnat}
 \bibliography{references}






\end{document}